\documentclass[5p]{elsarticle}

\usepackage{graphics}
\usepackage{lineno,hyperref}
\usepackage{graphicx}
\usepackage{amsfonts}
\usepackage{subfig}
\usepackage{ctable}
\usepackage[para,online,flushleft]{threeparttable}
\include{epsf}
\include{epstopdf}
\biboptions{longnamesfirst}
\DeclareGraphicsExtensions{.eps,.jpg,.pdf,.png,.tif}

\begin{document}
\newcommand{\beq}{\begin{equation}}
\newcommand{\eeq}{\end{equation}}
\def\la{\hbox{\raise.35ex\rlap{$<$}\lower.6ex\hbox{$\sim$}\ }}
\def\ga{\hbox{\raise.35ex\rlap{$>$}\lower.6ex\hbox{$\sim$}\ }}
\def\runit{\hat {\bf  r}}
\def\phunit{\hat {\bfo \bvarphi}}
\def\etaunit{\hat {\bfo \eta}}
\def\zunit{\hat {\bf z}}
\def\zetaunit{\hat {\bfo \zeta}}
\def\xiunit{\hat {\bfo \xi}}
\def\Kelvins{$\mathrm{K}$}
\def\beq{\begin{equation}}
\def\eeq{\end{equation}}
\def\beqa{\begin{eqnarray}}
\def\water{H$_2$O\ }
\def\CO2{CO$_2$}
\def\CO{CO\ }
\def\NH3{NH$_3$}
\def\CH4{CH$_4$}
\def\N2{N$_2$}
\def\eeqa{\end{eqnarray}}
\def\sub#1{_{_{#1}}}
\def\order#1{{\cal O}\left({#1}\right)}
\newcommand{\sfrac}[2]{\small \mbox{$\frac{#1}{#2}$}}
\begin{frontmatter}
\title{Modeling glacial flow on and onto Pluto's Sputnik Planitia\tnoteref{mytitlenote}}
\tnotetext[mytitlenote]{All place names of the Pluto system are informal.}

\author[mymainaddress,mysecondaryaddress]{O. M. Umurhan\corref{mycorrespondingauthor}}
\cortext[mycorrespondingauthor]{Main contact}
\ead{orkan.m.umurhan@nasa.gov}

\author[myterciaryaddress]{A. D. Howard}
\author[mymainaddress,mysecondaryaddress]{J. M. Moore}
\author[myMITaddress]{A. M. Earle}
\author[mymainaddress]{O. L. White}
\author[myquaternaryaddress]{P. M. Schenk}
\author[myMITaddress]{R. P. Binzel}
\author[SWRIaddress]{S.A. Stern}
\author[mymainaddress,mysecondaryaddress]{R.A. Beyer}
\author[UCSCaddress]{F. Nimmo}
\author[WashUaddress]{W.B. McKinnon}
\author[mymainaddress]{K. Ennico}
\author[SWRIaddress]{C.B. Olkin}
\author[JHUAPLaddress]{H. A. Weaver}
\author[SWRIaddress]{L. A. Young}

\address[mymainaddress]{National Aeronautics and Space Administration (NASA),
Ames Research Center,
Space Science Division,
Moffett Field, CA 94035
}
\address[mysecondaryaddress]{SETI Institute,
189 Bernardo Ave,
Suite 100,
Mountain View, CA 94043}

\address[myterciaryaddress]{
University of Virginia,
Department of Environmental Sciences,
P.O. Box 400123
Charlottesville, VA 22904-4123}

\address[myMITaddress]{
Massachusetts Institute of Technology,
Department of Earth, Atmospheric and Planetary Sciences,
Cambridge, Massachusetts 02139}

\address[myquaternaryaddress]{
Lunar and Planetary Institute,
3600 Bay Area Blvd.
Houston, TX 77058}

\address[SWRIaddress]{
Southwest Research Institute,
Boulder, CO 80302}

\address[UCSCaddress]{
Earth and Planetary Science,
University of California, Santa Cruz, CA 95064}

\address[WashUaddress]{
Department of Earth and Planetary Sciences, 
Washington University, 
St. Louis, MO 63130}

\address[JHUAPLaddress]{
Johns Hopkins University Applied Physics Laboratory, 
Laurel, MD, 20723
}

\begin{abstract}
Observations of Pluto's surface made by the New Horizons spacecraft indicate
present-day \N2 ice glaciation in and around the basin informally known as Sputnik Planitia.
Motivated by these observations,
we have developed an evolutionary
glacial flow model of solid \N2 ice that takes into account its published thermophysical and rheological properties.
This model assumes that glacial ice flows laminarly and has a low aspect ratio which permits a
vertically integrated mathematical formulation.
We assess the conditions for the validity of laminar \N2 ice motion by revisiting 
the problem of the onset of solid-state buoyant convection 
of \N2 ice for a variety of bottom thermal boundary conditions.
{  Subject to uncertainties in \N2 ice rheology, \N2 ice layers
are estimated to flow laminarly for thicknesses less than 400-1000 meters.}
The resulting mass-flux formulation for when the \N2 ice flows as a laminar dry glacier
 is characterized by an Arrhenius-Glen functional form.
The flow model developed is used here to qualitatively answer some questions motivated by
features we interpret to be a result of glacial flow found on Sputnik Planitia.  We find that the wavy transverse dark
 features found along the northern shoreline of Sputnik Planitia may be a transitory imprint of
 shallow topography just beneath the ice surface suggesting the possibility that a major
 shoreward flow event happened relatively recently, within the last few hundred years.  Model results also support the interpretation
 that the prominent darkened features
 resembling flow lobes observed
 along the eastern shoreline of the Sputnik Planitia basin may be the result of a basally wet
 \N2 glacier flowing into the basin from the pitted highlands of eastern Tombaugh Regio.
\end{abstract}

\begin{keyword}
\texttt{elsarticle.cls}\sep \LaTeX\sep Elsevier \sep template
\MSC[2010] 00-01\sep  99-00
\end{keyword}

\end{frontmatter}


\section{Introduction}
The New Horizons flyby of Pluto has revealed a planetary surface exhibiting
evidence of recent geological activity
\cite{Stern_etal_2015,Moore_etal_2016}. In a compendium study found in this volume
(Howard et al., 2017, this volume) we have examined the evidence for both 
ancient and recent glacial flow on Pluto based on imaging data
taken by the
MVIC and LORRI cameras
as well as spectroscopic data obtained by the LEISA instrument.
\footnote{These are cameras and instruments onboard New Horizons:
 MVIC (Multispectral Visible Imaging Camera)
 is the medium-resolution color wide-angle ``push-broom" camera
 while LORRI (LOng Range Reconnaissance Imager)
is the high resolution framing panchromatic camera.
LEISA (Linear Etalon Imaging Spectral Array) is
 the 256 channel imaging spectrometer  \cite{Stern_etal_2015,Reuter_etal_2008}.}
The aim of this work is to develop a physical model framework to use with investigating various scenarios pertaining to \N2 ice glacial flow into and within Sputnik Planitia (SP, hereafter 
{  and note that
all place names on Pluto are informal})
\footnote{An earlier version of this manuscript referred to this region as ``Sputnik Planum".  The designation
of this feature from {\it planum} to {\em planitia} is a result of recent topographic refinements indicating that this
planar region has an elevation below the mean radius of Pluto.}
.
\N2, CH$\sub4$ and CO are observed to be present within and around SP
\cite{Grundy_etal_2016}
as well as in Pluto's atmosphere
\cite{Gladstone_etal_2016}. 
Based on the relative overabundance of \N2 compared to CO on SP, we shall assume that 
it is the primary
material undergoing glacial flow -- however, this should be subject
to further scrutiny.
\par
\begin{figure*}
\begin{center}
\leavevmode
\includegraphics[width=16.cm]{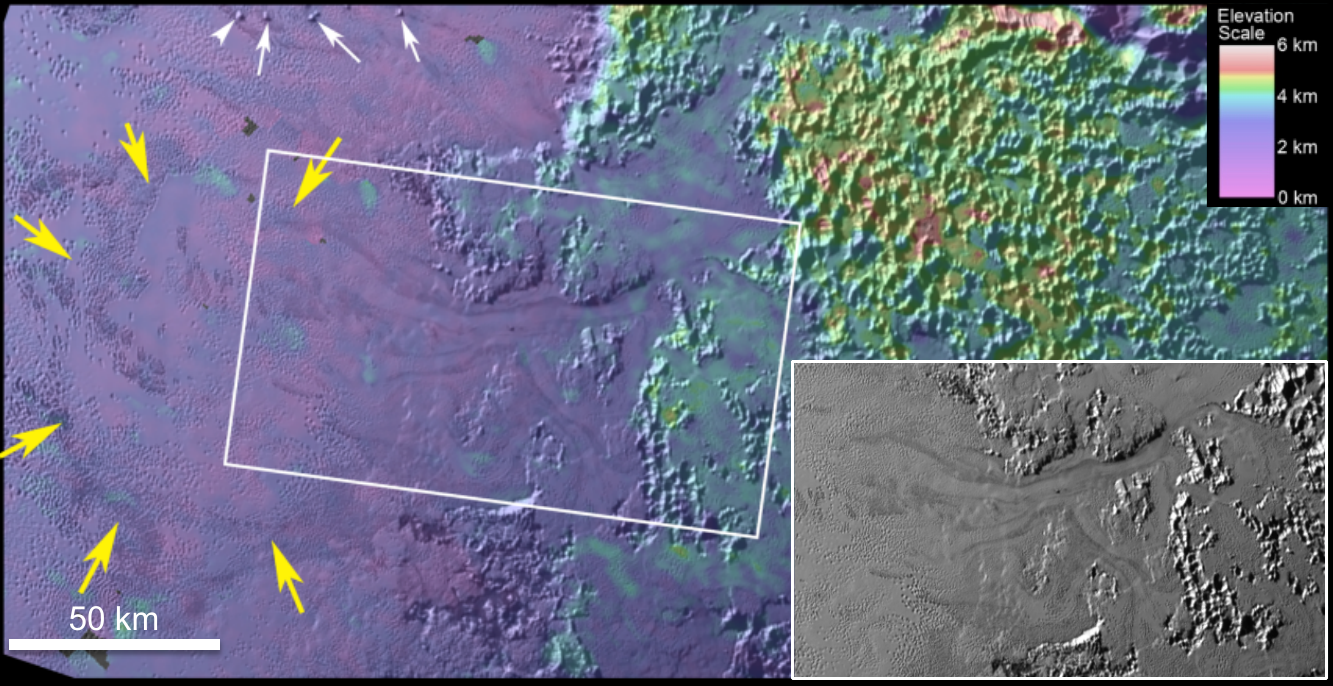}
\end{center}
\caption{A LORRI image section of eastern Sputnik Planitia abutting the pitted highlands
of ETR color-coded for elevation.  Streak patterns emanating from the higher elevations
are seen to follow an apparent flow path to the lower reaches of Sputnik Planitia.
Yellow arrows indicate the extent of flow lobes.
Examples indicating englaciated debris shown with white arrows.  In this image, north is up.
Location of this section is found
in Fig. \ref{LandScapeLocator}.
}
\label{Stunning_Glacial_Flow}
\end{figure*}

Figure \ref{Stunning_Glacial_Flow} depicts a prominent example of features suggesting
recent \N2 glacial flow from the pitted uplands of eastern Tombaugh Regio (ETR, hereafter)
and down onto the plains of SP. 
Dark streaks reminiscent of medial moraines, emanating from the highlands, connect to the floor of SP 
by commonly passing through
narrow throats 2-5 km wide.  The dark streaks terminate on SP with a pattern resembling flow lobes.
There are also lone protruding H$_2$O ice blocks with trailing dark streaks leading towards the higher
elevation \N2 icy flats, highly suggestive that these structures may be englaciated debris 
(Howard et al., 2017, this volume).  \par

\begin{figure*}
\begin{center}
\leavevmode
\includegraphics[width=8.57cm]{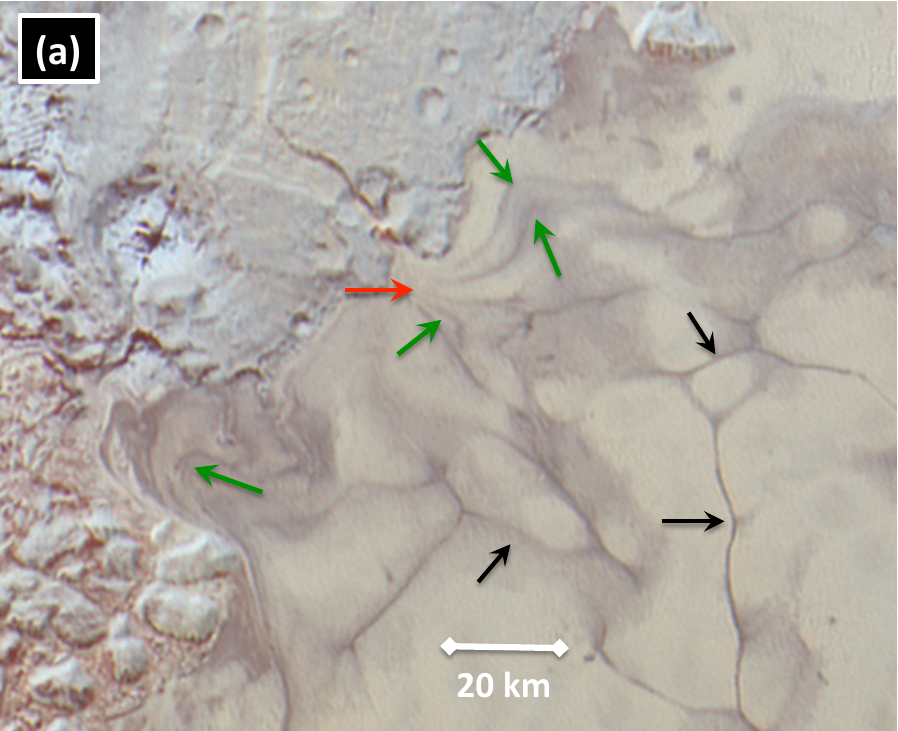}
\ \ 
\includegraphics[width=8.95cm]{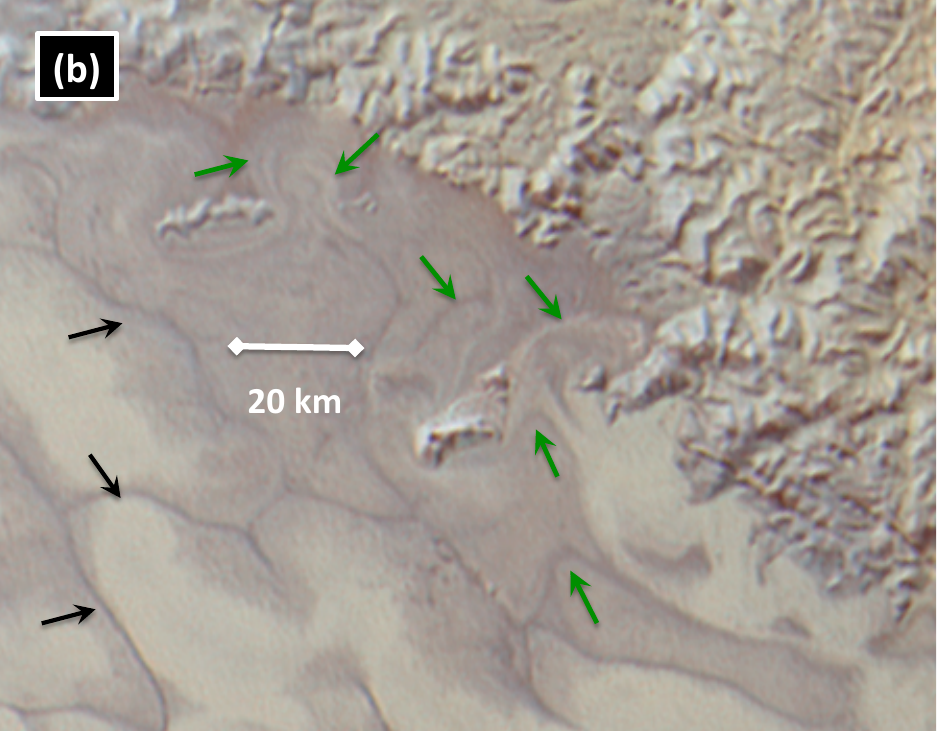}
\end{center}
\caption{MVIC sections of the northwestern (left panel) and northeastern (right panel) shorelines 
of Sputnik Planitia.
The dark ovoid patterning found on the lower portion of both images (black arrows)
have been interpreted as downwellings involving solid-state convection of \N2 ice
\cite{McKinnon_etal_2016}.
The near shoreline dark patterning (green arrows) may be indicative of either flow advection
of recently inactivated convective downwellings or an imprint of surface topography beneath the
observable surface. The red arrow indicates a possible undersurface drainage point (Howard et al. 2017, this volume). In both images, north is up.  Location of this section is found
in Fig. \ref{LandScapeLocator}.}
\label{Outflow_Image_1}
\end{figure*}

Interestingly, flow also appears to be moving from the interior of SP out toward both its northern
and southern shores (Howard et al. 2017, this volume).
 Figures \ref{Outflow_Image_1}(a-b) depict examples of indicators
 of flow toward Sputnik Planitia's northern shoreline.  The lower portion of 
 Figure \ref{Outflow_Image_1}(a)
 depicts ovoid
 patterning interpreted to be downwellings associated with
 solid-state convection of \N2 ice \cite{Stern_etal_2015,Moore_etal_2016,McKinnon_etal_2016}.
 However the near shoreline manifestation of the dark patterns takes on 
 a markedly different quality, becoming increasingly wavy closer to shore.  
 Figure \ref{Outflow_Image_1}(b) depicts similar features with the additional
 appearance of transverse surface patterning suggestive of viscous flow around an obstacle.
 Two ways in which these observations may be interpreted are: (i) the northward moving \N2 ice flow advects darkened, possibly
 inactive, convective downwelling zones giving rise to the observed near shoreline gentle undulations  
 or (ii) the wavy patterns are an imprint of topography beneath SP's surface.
 
 \par

In order to test the viability of scenarios pertinent to Pluto's
apparent present-day glacial flow, a physical model needs to be
constructed that takes into account, among other things, 
the known thermophysical and rheological
properties of solid \N2 and CO under the very cold physical
conditions of Pluto's surface.  Of the latter molecule, very little laboratory data is available
under Plutonian surface
conditions, but given CO's
similar molecular bonding structure to \N2, we assume that
its behavior resembles that of \N2 under these conditions \cite{Moore_etal_2016}. 
Most importantly, however, is the fact that
the rheology of solid \N2 has not been very well constrained to date (see discussion in section 2.2) and since many
of the results we report in this study depend upon the rheology, any future
developments and updates regarding solid \N2's rheology will likely
necessitate revising many of the results quoted here.  Despite these current
uncertainties, we can use this model to answer some broad qualitative questions
about the nature of the observed surface flow.
\par
According to experiments and analyses done to date (see Section 2.2 for further
details),
\N2 ice behaves as a viscous fluid with a viscosity somewhere
between \water  ice in slightly subfreezing terrestrial conditions ($\eta \sim 10^{12} \ {\rm Pa}\cdot {\rm s}$
at stresses of $10^5$ Pa \cite{Durham_etal_2010}),
and room temperature pitch or bitumen ($\eta \sim 10^{8} \ {\rm Pa}\cdot {\rm s}$).  
Conditions
affecting the flow of \N2 ice include the depth
of the flowing \N2 layers, the incoming geothermal flux 
and the temperature conditions both at the bases and
surfaces of such layers. The layer depth and surface temperatures
determine whether or not the \N2 ice moves as a ``basally wet", ``wet", or ``dry" glacier, and 
whether or not a glacier of a given thickness experiences buoyant solid state convection.
\footnote{A basally wet glacier here is understood to be a glacier whose base
is liquid.  This should be distinguished from a wet glacier which is, throughout its entirety,
 a mixture of ice and liquid between ice grains.}
A preliminary physical model should include other
various landform modification processes like 
accrescence/decresecence, i.e., continuous growth/retreat by accretion/decretion, \cite{Howard_Moore_2012}, and bedrock scouring. Of course, to what degree
these various landform modifying effects are operating on today's Pluto
is unclear, but we formally include it in our construction with the expectation
that these inputs will become clear once more research is done.
\par
With today's surface temperature on Pluto being $\sim$ 38.5 \Kelvins,  \cite{Grundy_etal_2016},
getting solid \N2 to surpass it triple point at the base of the layer requires layer thickness  
of up to 0.1-1km or more
\cite{Moore_etal_2016,McKinnon_etal_2016}.  
\N2 collected
on sloping channels will likely advect into depressions (including SP) long before one
can build up that much material owing to the apparent low viscosity
of solid \N2.  However,
occasional basal melt of \N2 ice layers
in the pitted highlands surrounding SP may have occurred in Pluto's recent geologic past.
Two recent studies, \cite{Earle_and_Binzel_2016} and (Earle et al., 2017, this volume),
have shown that Pluto's obliquity variations \cite{Dobrovolskis_Harris_1983}
coupled with the precession of Pluto's orbital apsis can give rise to 
conditions in which Pluto's surface temperatures might get as high as 
$\sim$ 55-60 \Kelvins \ for stretches of time lasting 2-3 decades.  Under those
conditions, \N2 ice layers as little as 200-400 meters thick may undergo basal melt
which could give rise to episodic, possibly flood-like, drainage events.
Similarly, however, basal melt can also result in steady-state drainage of liquid
\N2 into the \water ice bedrock below -- especially if the substrate is highly porous 
and/or supports a deeply extended crack system.
\par
At this early stage of discovery, these observations and reflections 
point to many questions about the history and conditions
of \N2 ices on Pluto.
In this study we start with two modest scenarios which we examine
using a vertically integrated physical model whose development is detailed in section 4.
The two questions we seek to address with our mathematical model are
\begin{enumerate}
\item Are the transverse wavy dark features near the northern edges
of SP an imprint of bottom topography?
\item Are the flow lobe features and deposits indicated by
similarly darkened features in Figure \ref{Stunning_Glacial_Flow}
evidence of basal melting and consequent basal sliding style flow?
\end{enumerate}
The short answer to each question is: (a) 
Our model results show that bottom topography can be embossed on the surface of flowing \N2
ice.  However, due to the viscous nature of the ice,
the imprint fades in a few decades suggesting that such features are
transitory and short.
(b)  A scenario in which \N2 ice flows as a basally wet glacier, 
{  i.e., a glacier whose base is in a liquid state},
better reproduces the quality
of the observed flow lobe features, however, it does not rule out that the observed features
are a result of purely dry glacial flow.

\par\medskip

This work is organized in the following way.  In section 2, we review known relevant information
on solid \N2 and CO, and we provide a table summarizing their thermophysical properties.  Section 2.2
provides a summary of the current knowledge of \N2 rheology including estimates of its
stress-strainrate relationships as well as the corresponding viscosities.  Section 2.3 discusses the near surface thermal profiles likely to be present within \N2 ice.  Section 2.4 considers the range of possible surface temperatures on Pluto over the course of its obliquity variations over the last 3 million years.
Section 3 examines the conditions for the onset of thermal convection in \N2 ice
heated from below by a static \water ice bedrock conducting
Pluto's geothermal heat-flux.  The answer to this question is important
because it determines the validity of the
glacial model developed in the section that follows.
\par
Section 4 contains a description of the flow model and mass-flux law we construct
for \N2 glacial flow appropriate for the surface of  Pluto
followed by
an examination of various e-folding relaxation timescales for \N2 ice structures of various
sizes, depths and extents using the formulated mass-flux law.
Owing to its highly technical nature, we have relegated to Appendix C the detailed
mathematical development
of the vertically integrated flow model we adopt which includes
the primary assumption that the flow is laminar, neither experiencing
buoyant convection nor shear flow induced turbulence.  
Section 5 
begins with detailing the model landscapes we use for our simulations and is concluded
by model results concerning the two
hypothetical questions posed above.  Section 6 reviews our efforts
and rounds out with some thoughts on future directions.

\section{General physical properties}
\subsection{Thermophysical properties of N$_2$ and CO}\label{thermophysical_properties}
\N2 and CO have similar molecular bond physics (primarily Van der Walls)
and their collective properties as a solid ought to be similar to one another.  However,
while some laboratory data is available for \N2 at the temperatures and pressures of relevance for
Pluto, comparatively scant laboratory data is available for CO.  
Nonetheless, we have compiled and summarized this data
in tabular form in Table \ref{N2_CO_properties_table}.  We note an important
quality of \N2, namely that it is relatively insulating compared to \water ice:
in fact the conductivity of nitrogen ice is $K({\rm N}_2) \approx (1/20) K({\rm H}_2{\rm O})$, 
where the figure for \water ice conductivity $K({\rm H}_2{\rm O})$ in the temperature range 
36-63 K is taken from reference \cite{Schmit_etal_1998}.  It is also noteworthy that the coefficient of thermal
expansion of nitrogen ice is also nearly 5 times that of \water ice at these temperatures.
\par
Both \N2 and CO have two solid phases: $\alpha$ and $\beta$.  The crystal structure
of the $\alpha$ phase is cubic while the $\beta$ phase
is a hexagonal close-packed crystal \cite{Scott_1976,Fracassi_etal_1986}.
The $\alpha-\beta$ phase transition in CO occurs at $T = T_{_{\alpha\beta}}({\rm CO}) = 61.55\  $K  while
the corresponding transition occurs for \N2 at $T = T_{_{\alpha\beta}}({\rm N}_2) = 35.6\ $K
\cite{Fray_Schmitt_2009}
with transition enthalpies of 22.4 kJ/kg and 8.2 kJ/kg for CO and \N2 respectively
\cite{Fracassi_etal_1986,Giauque_Clayton_1933}.
We note that the surface pressures are sufficiently low as to not effect
the quality of the emerging crystal structure \cite{Scott_1976}.
It is also worth noting that for the surmised surface temperatures
of Pluto, CO is in its $\alpha$-phase while \N2 is in its $\beta$ phase.
{  While CO and \N2 are fully soluble in the range of temperatures of interest to us,
the rheological properties of a composite mixture of the two molecular species is
currently unknown, requiring detailed laboratory work in the near future.
The uncertainty in the rheological properties are likely compounded when a
third constituent, like \CH4, is added to the mix.}

\begin{table*}  
\caption{Known properties of solid \N2 and CO near their triple points}
  \label{N2_CO_properties_table}
  \begin{center}
  \begin{threeparttable}
  \begin{tabular}{cccccccc}
    \toprule
 \ Molecule  \ & 
 \ \ $T_m/P_m$\tnote{a}  \ \ &  
\ \ \ \ \ $\rho_s$  \ \ \ &  $C_p$ & $K$ & $\alpha$ \tnote{b}
&\ \ \ $\Delta E_s$\tnote{c} \ \ \  & \ \ \ $\Delta E_f$\tnote{d}\ \ \ \\ 
 & (\Kelvins/bars) & (g/cm$^3$) & \ \ \ (kJ/kg\,$\cdot$\Kelvins) \ \ \ & 
 \ \ \ (W/m\Kelvins) \ \ \ & ($\times 10^{-3}/ $\Kelvins)
 & \ \ \ (kJ/kg) \ \ \   & \ \ \ (kJ/kg) \ \ \  \\ 
    \hline
    \hline
\N2 & 63.10/0.11 \tnote{e} & 0.99-0.95 \tnote{g} & 1.35-1.65 \tnote{g} & 0.215-0.200 \tnote{g} & 1.7-2.7 \tnote{g}
 & 224.9 \tnote{h} & 25.7 \tnote{h}  \\ 
CO & 68.13/0.14 \tnote{i} & 1.01-0.97 \tnote{j}  & 1.29-2.32 \tnote{k,l} & 0.350-0.280 \tnote{m} & n/a & 296.4 \tnote{l} & 29.6 \tnote{n}
 \\ 
\bottomrule
\end{tabular}
\begin{tablenotes}
\footnotesize
\item[a] Melt temperature and pressures at triple point.
\item[b] Coefficient of thermal expansion.
\item[c] Heat of sublimation in vicinity of $T_m$.
\item[d] Heat of fusion in vicinity of $T_m$.
\item[e] Ref. \cite{Fray_Schmitt_2009}.
\item[g] Range quoted for $\beta$-phase of \N2 between 36-63 \Kelvins, ref. \cite{Scott_1976}.
\item[h] Ref \cite{Giauque_Clayton_1933}.
\item[i] See \cite{Fray_Schmitt_2009,Goodwin_1985}
\item[j] Range quoted for $\alpha$-phase of CO between 41-61.5 \Kelvins, based on compiled
data shown in Fig. 1 of ref. \cite{Fracassi_etal_1986}.
\item[k]  Note that in $\alpha$-phase these quantities deviate from linear profile.
\item[l]  Experimentally determined, ref. \cite{Fracassi_etal_1986,Kelley_1935}
\item[m]  Range quoted for $\alpha$-phase of CO between 45-65 \Kelvins,
based on experimental data shown in Fig. 4 of Ref. \cite{Konstantinov_etal_2006}. 
\item[n]  Ref. \cite{Clayton_Giauque_1932}.

\end{tablenotes}
\end{threeparttable}
\end{center}
\end{table*}

      \begin{figure}
\begin{center}
\leavevmode
\includegraphics[width=8.5cm]{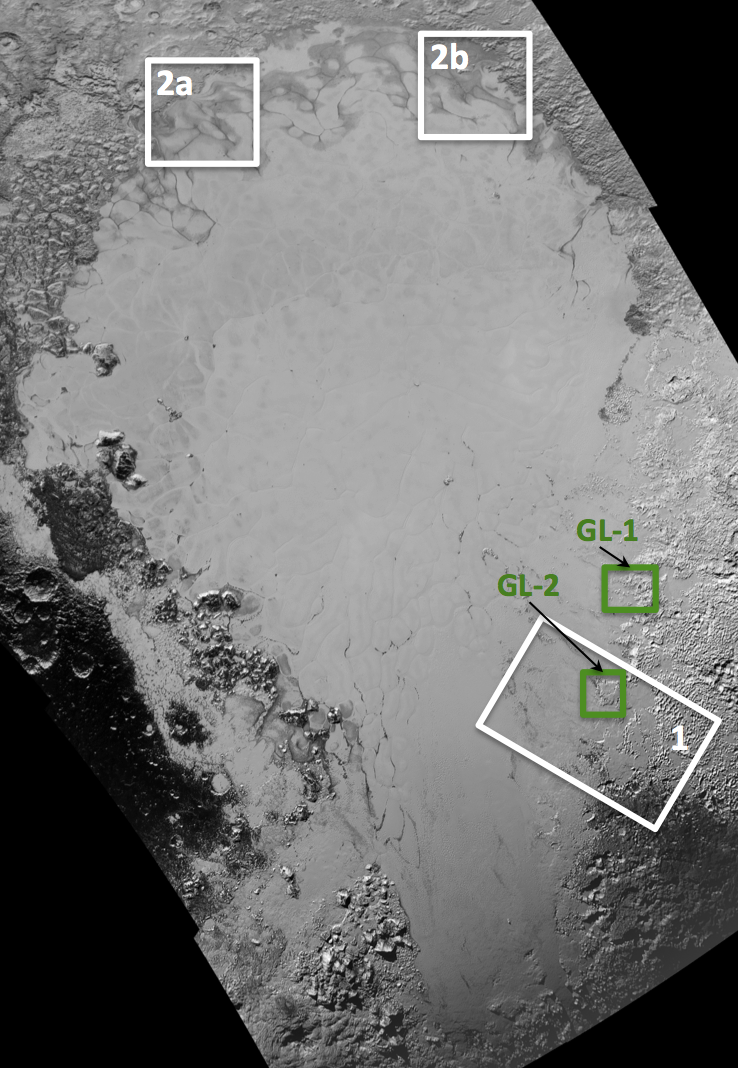}
\end{center}
\caption{A larger view of Sputnik Planitia with the locations of
Figures 1-2 indicated (white boxes).  The green boxes
indicate sample landscapes (``GL-1", ``GL-2") are used
as test bedrock landforms for modeling done in section 5.
}
\label{LandScapeLocator}
\end{figure}

\subsection{Rheological properties of N$_2$ and CO}\label{rheological_properties}
As of the writing of this work, there are two known studies pertaining
to the rheological properties of \N2 in solid-state for the temperatures
of interest to us, i.e. between 30 K and 63 K.  There is no known
analogous laboratory analysis of CO.
\footnote{Probably owing to its hazardous biological toxicity.}
  As such, and  primarily
owing to the similar
molecular bond structures of both \N2 and CO (see last section), we cautiously consider
the qualitative and quantitative properties of CO ices to be
similar to \N2.  We briefly review the aforementioned
\N2 studies here:
\par
\bigskip
{\it{Non-newtonian creep of laboratory annealed solid \N2}}.
 \cite{Yamashita_etal_2010}
 consider the rheological properties of laboratory annealed \N2 under 
 a variety of tangential stresses and at two temperatures (45 K and 56 K).  They show that
 its stress, $\sigma$, and strain-rate, $\dot\epsilon$, behave with a power-law
 relationship given by
 \beq
 \dot\epsilon = A(T) \sigma^n(T),
 \eeq
 where the prefactor $A$ and the index $n$ (in the vicinity of 2.1) 
 are both temperature dependent.
 The prefactor $A(T)$ was fit to an Arrhenius form in
 \cite{Moore_etal_2016} and shown to have an effective activation energy $E_a$
 of about 3.5 kJ/mole or, equivalently
 \beqa
 & & A(T) = A\sub{45} 
 \exp\left({{\frac{T_a}{45 \ K}-\frac{T_a}{T}}}\right), 
 \nonumber \\
 & &  A\sub{45} = 0.005 s^{-1} ({\rm MPa})^{-n(T)},
 \label{Glen_Prefactor_N2}
 \eeqa
 where the activation temperature is $T_a= E_a/k = 422 K$. 
 The temperature
 dependence of the index was determined assuming a linear
 fit to be given by the expression  $n(T) = 2.1 + 0.0155(T/K - 45)$.  With this
 in mind, we assume a {\it visco-plastic viscosity} which is defined as
 \beq
 \eta \equiv \frac{\partial\sigma}{\partial\dot\epsilon},
 \label{def_solid_state_viscosity}
 \eeq
 \cite{Hughes_2009}.  Together with this we input the power-law form for the laboratory annealed \N2 ice experiment
 to find
 \beq  \eta = \frac{1}{n A \sigma^{n-1}}.
 \eeq
 Under conditions where the tangential stresses are about 10 kPa (0.1 bar)
 the viscosity of \N2 ice at 45 \Kelvins \ is about $1.6\times 10^{10}$ Pa$\cdot$s.
 

\par
\bigskip
{\it{Newtonian creep of \N2 ice grains}}.
In \cite{Kirk_1990,Eluszkiewicz_Stevenson_1990,Eluszkiewicz_1991} the
rheological properties of \N2
{\emph{ice grains}}
were examined in the framework of both Coble creep and Nabarro-Herring (NH) creep, both considered
diffusional processes ($n=1$)
and often considered collectively as grain-boundary sliding (GBS) \cite{Durham_etal_2010}.  As opposed to the non-Newtonian case just discussed,
we assume the viscosity of grains is linear (and, hence, ``Newtonian").
The grain rheologies discussed in these aformentioned studies are based on
a nuclear magnetic resonance (NMR) study of \N2 grains 
done by \cite{Esteve_and_Sullivan_1981}.  Coble creep is based on the
diffusion of molecules {\emph{along}} grain boundaries while NH creep is based on
the diffusion of molecules {\emph{through}} grains themselves.  The general trend
is that NH-creep dominates Coble creep when a material approaches its melt temperatures.
Based on the scant data available for \N2, NH creep will dominate Coble creep
for temperatures greater than about 25K for grain diameters in
the vicinity of 1mm (see \ref{relative_creep_strengths}).
\par
Thus, for \N2 ice grain spheres of radial
size $d_g$ the effective volume diffusion rate through grains
$D_{0v}$ was empirically 
assessed by \cite{Esteve_and_Sullivan_1981} (studying the self-diffusion
of \N2 ice laced with Ar impurities) 
to be given by
\beq
D_{0v} = 1.6 \times 10^{-7} e^{-T_v/T} \ {\rm m^2/s},
\eeq
where the experimentally determined activation temperature $T_v \approx 1030$K
relates to the corresponding experimentally determined
activation energy, $E_v \approx 8.6 kJ/mole$,
via $E_v \equiv k T_v$, where $k$ is the Boltzmann constant
($=1.38\times 10^{-23}$m$^2$kg\,s$^{-2}$K$^{-1}$).  From this relationship
an effective {\it linear viscosity} can be written as \cite{Eluszkiewicz_Stevenson_1990,Eluszkiewicz_1991}
\beq
\eta = \frac{kT d_g^2}{42 D_{0v} \Omega},
\eeq
where $\Omega$ is the volume of a single \N2 molecule ($=4.9\times10^{-29}$m$^3$).  This expression can
be rewritten in a way more revealing and relevant to  the surface of Pluto
\beq
\eta_{_{nh}} = 1.84 \times 10^{16} \left(\frac{d_g}{{\rm mm}}\right)^2 \left(\frac{T}{45\ K}\right)
\exp\left(\frac{T_v}{T}   - \frac{T_v}{45\ K} \right),
\label{eta_nh}
\eeq
in units of Pa$\ \cdot$ s.  
{ 
We have designated this effective viscosity
$\eta_{_{nh}}$ in order to make it clear that this is the purely ``Newtonian", diffusion creep 
instance of this form.   \par
While the non-Newtonian properties of \N2 ice grains are not
fully explored beyond that described above, we posit, by analogy to extensive studies
done on the creep properties of \water ice,  that diffusion GBS creep of \N2 ice grains
may be enhanced by dislocation accommodated GBS.  Assuming that the 
stresses in subsurface Plutonian ices sit well within the range of 1 - 10$^5$ Pa
and the grain sizes are within 0.001 to 1 mm, and accepting the posited
analogy between \N2 and \water \ ices, an
examination
of the various expected rheological properties of \water ices
showcased in Figure 6 of \cite{Durham_etal_2010}
suggests that \N2 ice
grains might have a ``superplastic" weakly non-Newtonian 
flavor with a stress-strain power-law index of $n=1.8$.
Provided this is the case for \N2 ice grains, then according to the 
definition of viscosity in Eq. (\ref{def_solid_state_viscosity})
we suggest that their collective viscous behavior might be described
by
\beq
\eta = \eta_{_{nh}} \left(\frac{\dot\epsilon_{_2}^{(nh)}}{\dot\epsilon_{_2}}\right)^{(1-n)/n}
=
\eta_{_{nh}} \left(\frac{\dot\epsilon_{_2}^{(nh)}}{\dot\epsilon_{_2}}\right)^{-4/9}
,
\label{proposed_N2_viscosity}
\eeq
where $\dot\epsilon_{_2}^{(nh)}$ is a reference value
of the second invariant of the strain tensor, a figure currently unknown and
requiring future laboratory work to determine.}

\subsection{On the thermal profiles of Plutonian \N2/CO ice layers}\label{thermal_properties}
The thermal gradients and timescales of \N2/CO ice-layers are
important to assess.  Assuming ice-layers too shallow for
convection to occur, we can estimate the vertical temperature gradient across a layer
of depth $H$.
A recent study by \cite{Robuchon_Nimmo_2011}, investigating the interior thermal history of Pluto,
suggests that the current-day geothermal heat-flux amounts to
a few $10^{-3} W/m^2$.  We adopt as a reference value for this
interior flux of $F^{\rm{\small (ref)}}_{_{{\rm Pl}}} = 4 \times 10^{-3} W/m^2$.
From the numbers compiled in Section \ref{thermophysical_properties} for \N2
\footnote{Keeping in mind that all quoted quantities have a temperature dependence
but we consider them sufficiently weak for our current considerations.}
, we adopt
a value for the conductivity $K \approx 0.2$ W/mK and assigning $F_{_{{\rm Pl}}}$ to be the emergent heat
flux, we find that the mean interior temperature gradient $\overline T_z
\equiv F_{_{{\rm Pl}}}/K $, or
\beq
\overline T_z \approx 20 \Big(F_{_{{\rm Pl}}}/F^{\rm{\small (ref)}}_{_{{\rm Pl}}}\Big)
\ \ ^\circ {\rm K}/{\rm km}.
\label{temperature_gradient_in_Pluto_N2_ice}
\eeq
We note that due to its insulating nature compared to \water \ ice, the vertical temperature
gradient in such ices is very high.
The thermal diffusion coefficient through solid \N2 can be estimated from the definition
$\kappa \equiv K/\rho_s C_p \approx 0.66 \times 10^{-6} {\rm m}^2/{\rm s}$
where
we adopt a value for the heat capacity $C_p
= 1.65 {\rm kJ}/{\rm kg} \ K$.  
Thus, one can estimate
the e-folding thermal relaxation time across a layer of width $H$ by $\tau \equiv H^2/(4\pi^2 \kappa)$,
which we express as \citep{Moore_etal_2016},
\beq
\tau = 6\times 10^{-3} (H/m)^2 \ \oplus{\rm yr} = (H/H_0)^2 \ \oplus{\rm yr},
\eeq
in which $\oplus\ {\rm yr}$ is one terrestrial year. The reference scale $H_0 \approx 12.9 \ $m, which says that a layer of approximately 13 meters
will equilibrate into its conductive profile in one Earth year.  
Thus, unless deposition rates greatly exceed this figure, the temperature profile of the 
glacier should be roughly equal to the conductive equilibrium profile.
Calculations performed and quoted by \cite{Young_2012} and \cite{Grundy_etal_2016}   
estimate sublimation/deposition rates on present-day Pluto
to be in the range 1-10 cm/yr which suggests that it is reasonable
to assume a glacier whose interior temperature is in conductive equilibrium.
%

\begin{figure*}
\begin{center}
\leavevmode
\includegraphics[width=8.2cm]{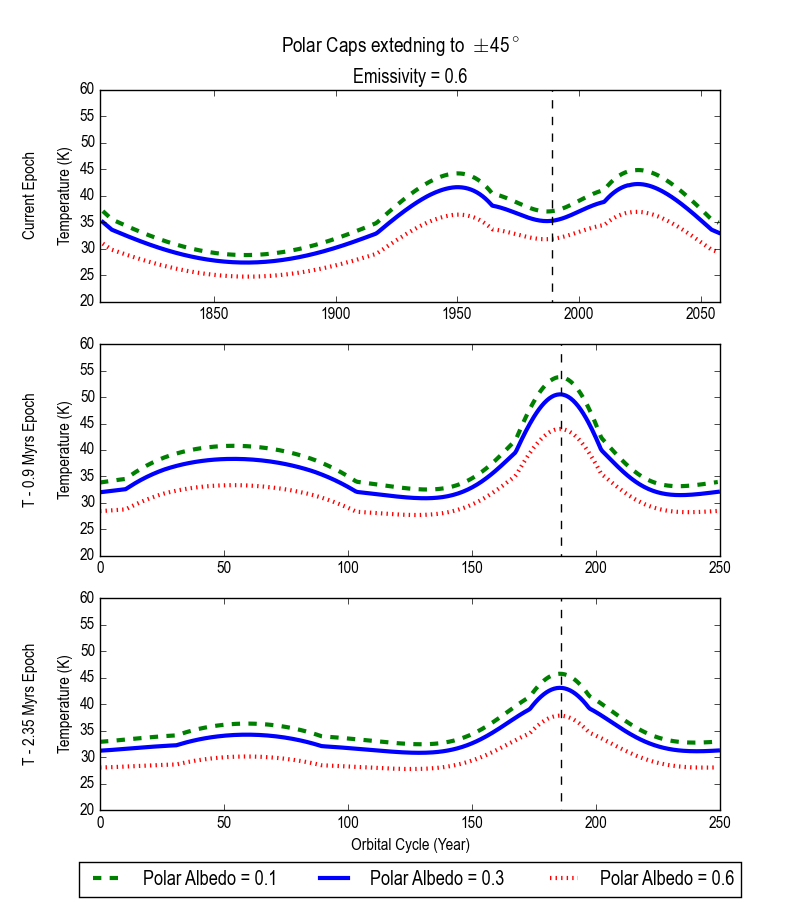}
\qquad
\includegraphics[width=8.2cm]{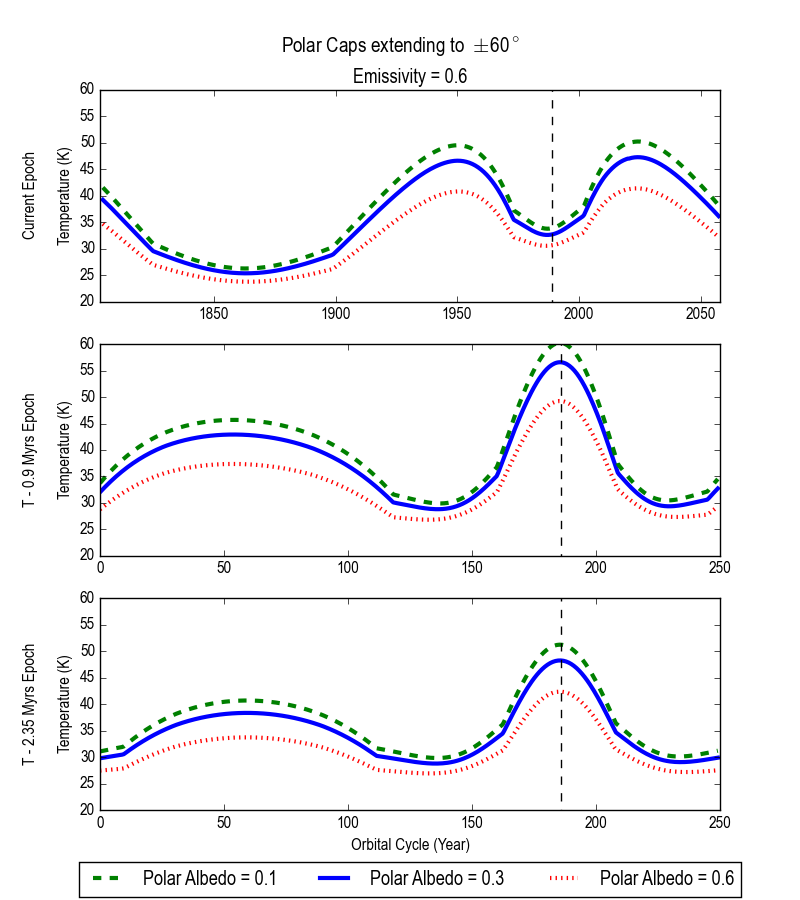}
\end{center}
\caption{Surface temperatures of Pluto over a 3 Myr obliquity/precessional
cycle.  Model assumes the presence of a landmass representing Tombaugh Regio
(see text).  The ice emissivity is assumed to be 0.6.  Solutions are
shown for three different values of the polar albedos.  Left panel shows a suite of
solutions assuming that the polar caps extend down to
$\pm ^\circ $45 latitude while right panel 
shows solutions with the
polar caps extended further to $\pm ^\circ $60 latitude.
 }
\label{Cap45Em6}
\end{figure*}


\subsection{Surface temperatures over time}\label{Pluto_Surface_Temps_Over_Time}
{  We undertake a momentary detour in our discussion and consider the past
history of Pluto's surface temperatures.  The import of this consideration
is to give a justification for the potential of \N2 ice glaciers to undergo
basal melt for certain stretches of time (see Section 4). 
The idea we have in mind here is the following:  with the emerging geothermal 
flux held fixed, a rise in the surface temperature will steadily induce a concomitant 
rise in the temperature of the base.  Thus, even if the surface does not approach melting point, the base of the glacier will more easily achieve this
state owing to its relative warmth.
 With a wet base, the flow
and evolution properties of \N2 glaciers will exhibit significantly different
qualities from those glaciers that remain basally frozen.}
\par
While
the surface temperature of today's Pluto is just under 40 
\Kelvins, the surface was likely to be much warmer in its past.
 Pluto's obliquity variations \cite{Dobrovolskis_Harris_1983}
coupled with the precession of its orbital apsides implies
that the total amount of seasonally averaged radiant energy received on a given hemispheric surface will significantly
vary over timescales of approximately three million $\oplus$ years.  This has direct influence
on the resulting surface pressures and temperatures \cite{Earle_and_Binzel_2016}
and can give rise to surface temperatures approaching 55-60 \Kelvins \  lasting for
about 20-30 Pluto years under conditions of favorable alignment of Pluto's orbital apsis
and obliquity extremes.
\par
Figures \ref{Cap45Em6}a-b show the surface temperature results of 
two Pluto climate models whose details may be found in Earle et al. (2017, This Volume).
The model, based on the one constructed by \cite{Trafton_1984}, and
its parameter inputs are summarized here:  Pluto's heliocentric distance and sub-solar latitudes
are calculated as a function of time based on the calculation of \cite{Dobrovolskis_etal_1997}.
A simplified albedo and volatile map is used
which assumes polar caps with latitudinal extent of either $\pm ^\circ $45 (Fig. \ref{Cap45Em6}) or 
$\pm ^\circ $60 (Fig. \ref{Cap45Em6}b)
.  
The polar caps are modeled as having a variety of different albedos (0.1, 0.3, or 0.6).
To represent the thermal influence of Tombaugh Regio,
Earle et al. (2017, this volume)
include a bright (albedo=0.6) volatile patch extending from $-^\circ$30 degrees 
to +$^\circ$45 degrees with a longitudinal extent of $^\circ$45. 
For different trials,
two extremes of a range of plausible emissivity values are examined (e.g. 0.6 or 0.9).  
These inputs are then incorporated into 
the energy balance equation model, originally presented in \cite{Trafton_1984},
and used to determine the global temperature as a function of time.
Note that the solutions developed do not take into account the possibility
of time lags associated with thermal inertia effects.  
\par
The results are fairly
typical: During extreme insolation conditions when Pluto's orbital apsides and obliquity are
favorably aligned, maximum surface temperatures persist for
20-30 Pluto years. 
 When the polar caps extend to $\pm ^\circ $60 latitude, the surface temperatures
become very close to the melt temperature of \N2.  This means that one may have circumstances in which
{\emph{relatively shallow}}
layers of \N2 ice may experience basal melting for stretches of time and this, in turn, 
can have important 
physical consequences
in the flow response of \N2 ice layers, i.e., whether
or not the layer exhibits basally wet or dry flow -- see further
discussion of this in section \ref{modeling_framework}.
\par
For example,
during these climate extremes, the surface temperatures get as high as $55$K which,
assuming a conductive temperature gradient within the ice of $20$ K/km, would
correspond to a layer thickness just under $400$ m before melt occurs at the base of the glacier
(the minimum pressure requirement for melt, of about $10^4$ Pa \cite{Fray_Schmitt_2009}, is satisfied at that depth).


\section{Onset of convection}\label{convective_onset}

The eastern shoreline of SP, and various locations along the other
boundaries of SP, indicate there are regions in which convection is either absent
or occurring very weakly \cite{McKinnon_etal_2016}.  While the subsurface basin profile of SP is unknown,
estimates as to the depth of various sections of SP can be made
based on the distinctive lack of obvious signs of buoyant solid-state convection.
One of the important inputs for this study is to estimate an upper limit of the thickness of 
an \N2/CO ice layer before it starts to undergo solid-state convection.  
In addition to knowing when one ought to expect convection to occur in, SP
the glacial flow modeling detailed in Section \ref{glacial_flow_modeling}
also implicitly assumes that the flowing glacial ice is more or less laminar,
which we assume to be the case if the layer thicknesses are less than
that for onset of convection.  It is therefore important to get a good
sense of the conditions under which convection is expected to occur. 
Thermal convection in terrestrial ice glaciers has been examined
as a possible explanation for the origin of ice streams which
can drain up to 90\% of glacial ice \cite{Hughes_2009}.
\par

\begin{figure}
\begin{center}
\leavevmode
\includegraphics[width=8.7cm]{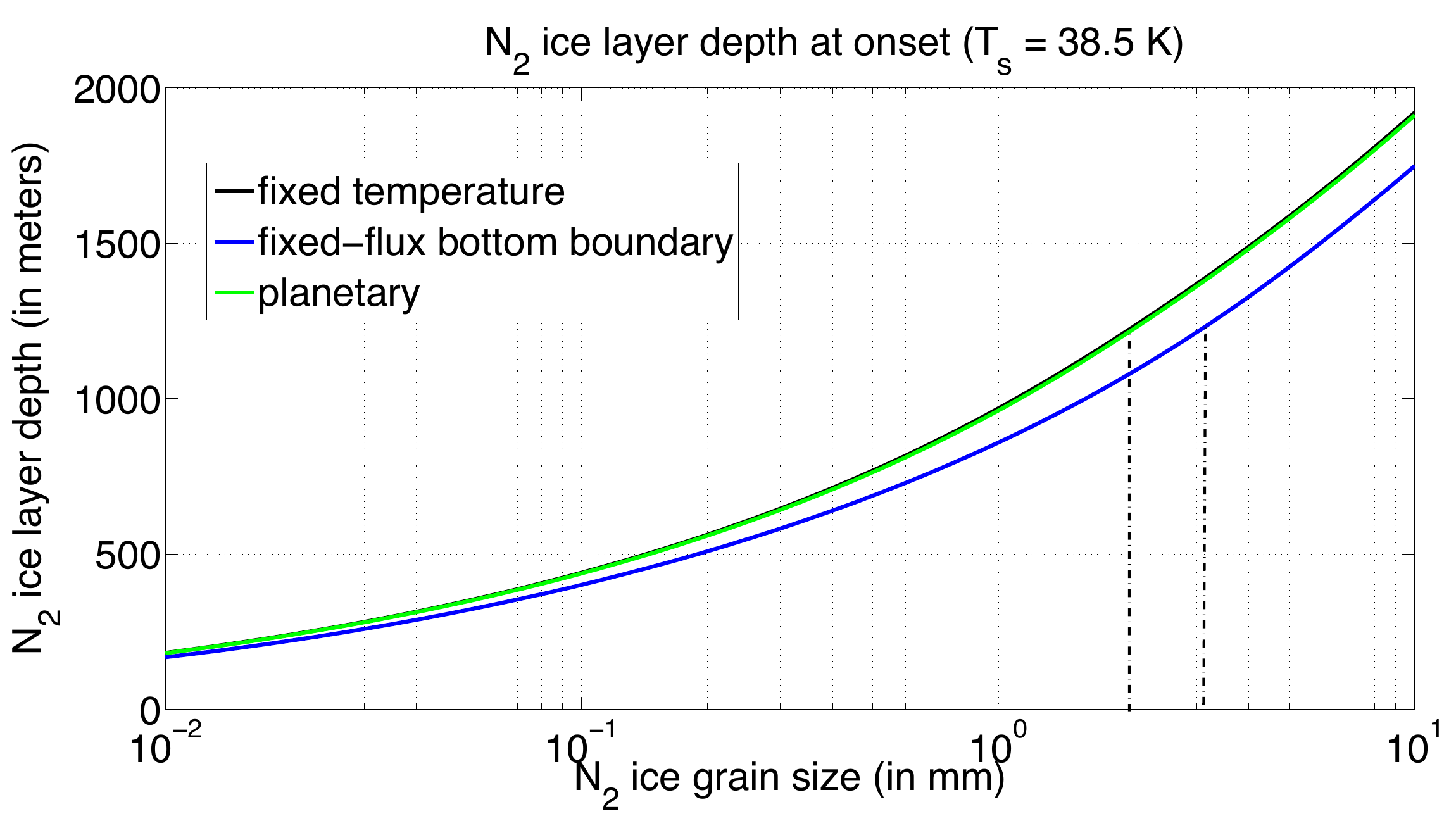}
\end{center}
\caption{Minimum layer thickness for the onset of convection as a function of \N2
ice-grain size.  Figure shows solution of transition to convection 
 assuming the top of the convecting layer has fixed upper temperature of
$T = 38.5$\Kelvins \ and vertical temperature gradient $\overline T_{_{0z}} = 20$ \Kelvins/km. 
Solutions for three different bottom thermal boundary conditions
shown: fixed-temperature, fixed-flux and ``planetary" -- detailed in \ \ref{onset_to_convection}.
 Solutions determined use the Frank-Kamenskii approximation \cite{Solomatov_2012}.  Vertical
 hatched lines denote critical grain-sizes for which
 the temperature at the layer's base corresponds to $T_m = 63.1$ \Kelvins,
 i.e., the triple-point melt temperature for \N2.
 }
\label{Min_Layer_thickness_for_Convection}
\end{figure}

\begin{figure*}
\begin{center}
\leavevmode
\includegraphics[width=8.7cm]{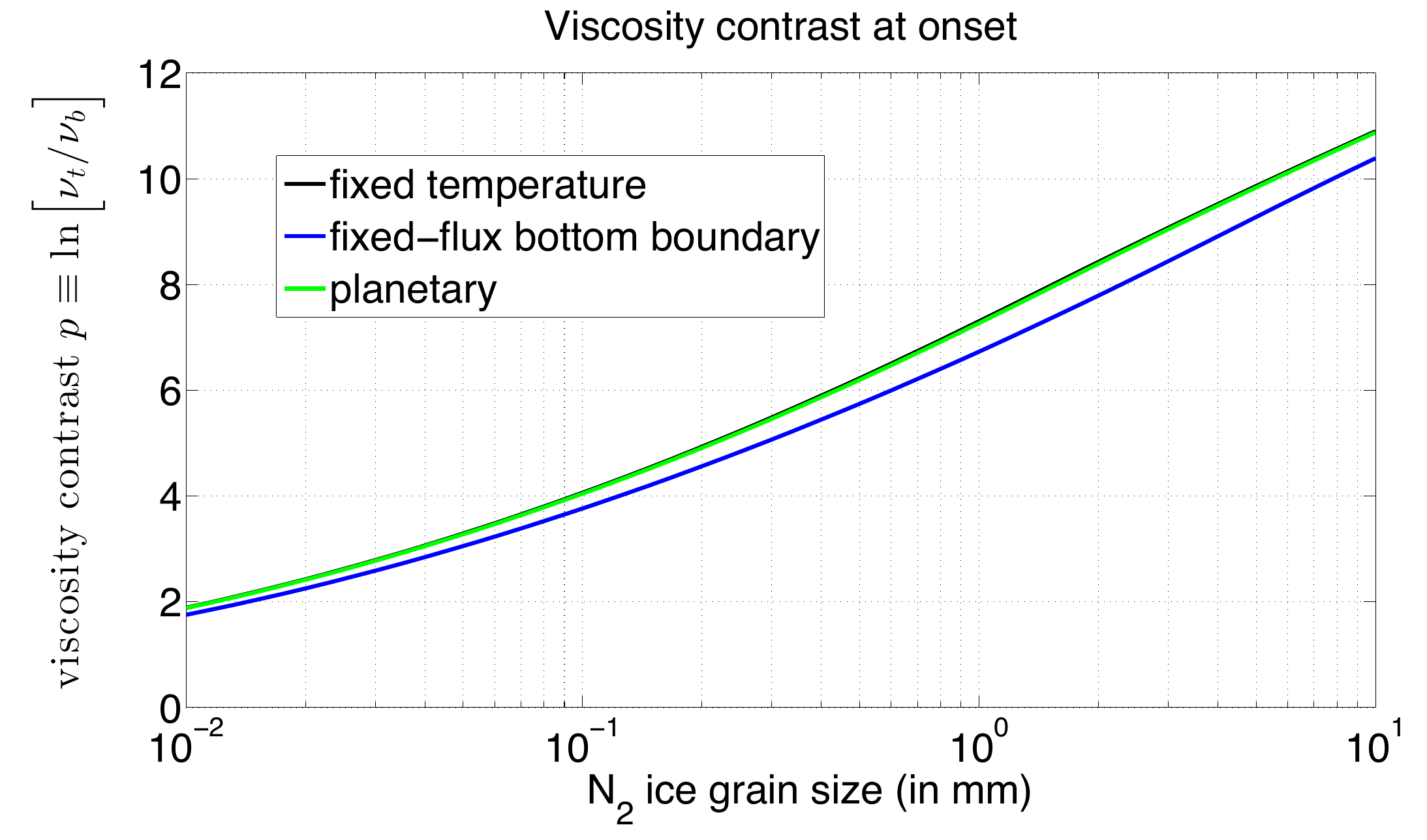}
\includegraphics[width=8.7cm]{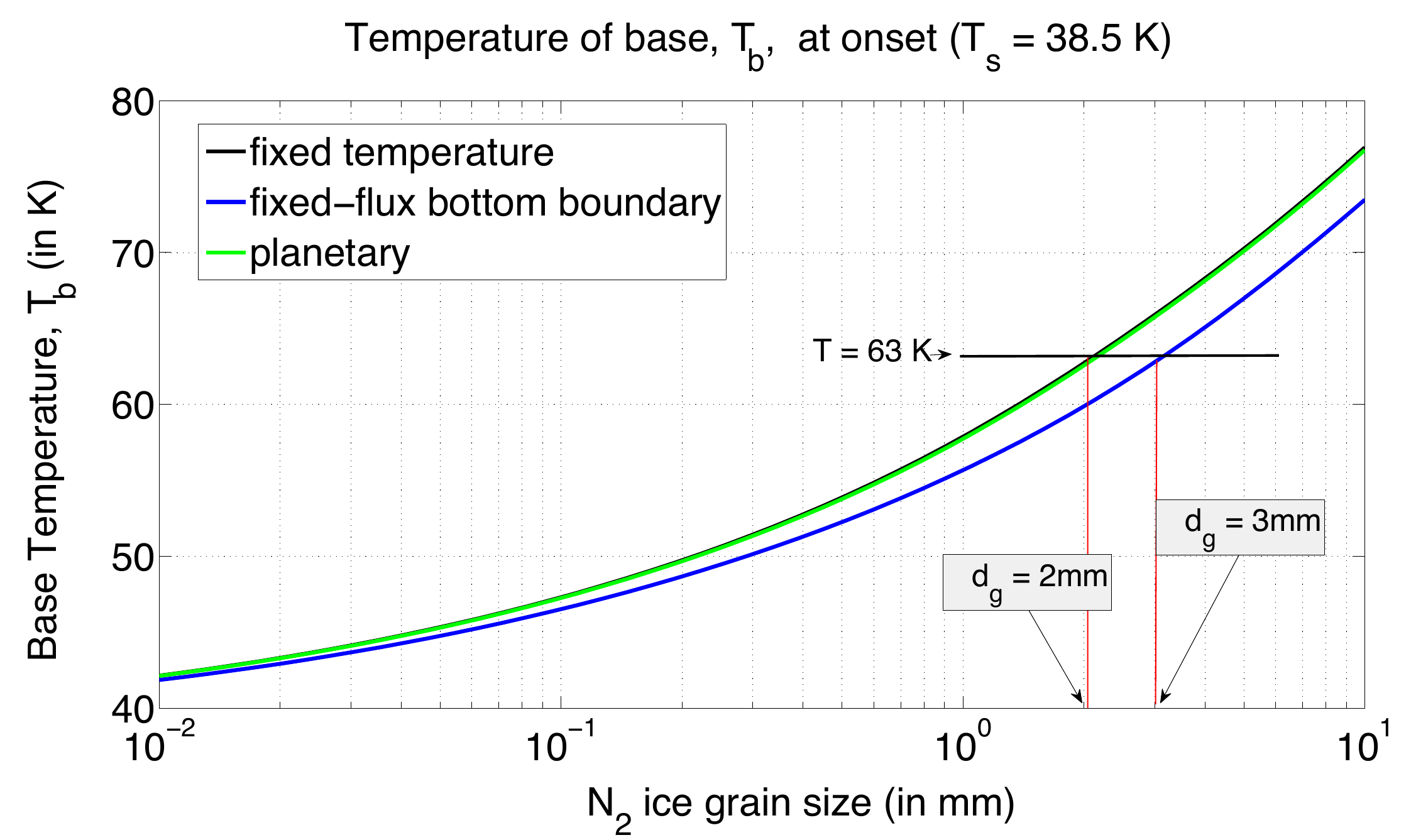}
\end{center}
\caption{(Left panel) Base temperature at transition
into buoyant convection.
(Right panel) Viscosity contrast for the onset of convection.
Results shown in both panels assessed for parameters shown in Fig. \ref{Min_Layer_thickness_for_Convection}.
The grain-size values corresponding to melt temperatures 
at the triple-point for \N2 are designated $d_g \approx 2$mm
for fixed-temperature and planetary thermal boundary conditions
at the base
while $d_g \approx 3$mm for fixed-flux bottom boundaries.
}
\label{Viscosity_contrast_for_Convection}
\end{figure*}

\par
It is not useful to perform here an exhaustive examination of
the transition into convection given all the possible formulations
of \N2's rheology outlined in Section \ref{rheological_properties}.
We are interested, however, in getting a good order of magnitude
figure for this onset. Thus
for our calculation here we assume a static Newtonian layer of \N2 ice grains
and consider the onset to convection assuming the viscosity is
given by $\eta = \eta_{_{\rm nh}}$, as defined in Equation (\ref{eta_nh}).
 This analysis then becomes amenable to a linear stability calculation.
We assume the layer is in conductive thermal equilibrium with its
interior which is a reasonable
assumption given the thermal equilibration times we estimated in 
the previous section.  The vertical horizontally uniform
temperature profile is $T(z) = T_s - \overline T_{_{0z}}(z-z_s)$
where $\overline T_{_{0z}}$ is given in Eq. (\ref{temperature_gradient_in_Pluto_N2_ice}),
 $z_s$ is the location of the surface, and $T_s$ is the temperature of the
 surface for which we set as the present day surface temperature
 on \N2 ice of 38.5 \Kelvins.
\par
Due to the temperature contrast, the viscosity at the top and bottom of a layer will
be different and, therefore, we consider the Rayleigh number Ra as
measured at the base of a given layer.
We hereafter consider the viscous diffusion $\nu \equiv \eta/\rho_s$ and
define define $p \equiv \ln(\nu_t/\nu_b)$ to be the natural
logarithm of the ratio of the viscosities between the top and bottom
layers, $\nu_t$ and $\nu_b$ respectively.  
The Rayleigh number Ra at the base is therefore defined as
\beq
Ra \equiv \frac{g\alpha H^4 \overline T_{0z}}{\kappa \nu_b},
\eeq
where $g = 0.642 m/s^2$ is the surface gravity of Pluto and $\alpha$
is the coefficient of thermal expansion whose values of \N2 ice
are given in Table \ref{N2_CO_properties_table}.
The onset to convection is a function of the two parameters $p$ and Ra.
We note that Ra has an Arrhenius law dependence due to the temperature
dependence on $\nu_b$, as well as a $d_g^{-2}$ dependence on ice-grain size.
The detailed form of Ra, as well
as the assumptions and solution methods employed to assess
this transition to convection, is detailed in \ref{onset_to_convection}.
\par
We are interested here in three questions: (1) For a given \N2 ice-grain size
$d_g$, how deep a layer of \N2 ice on Pluto does one need to initiate 
solid-state convection?
(2) At onset of solid-state convection, what is the critical viscosity
contrast as a function of $d_g$?  (3)
What is the temperature of the base of the layer
at onset to convection as a function of $d_g$?
  The answer to these questions are
graphically displayed in Figures 
\ref{Min_Layer_thickness_for_Convection}-\ref{Viscosity_contrast_for_Convection}
and we use these figures for our future referencing henceforth.  We note
that the depth of the ice-layer must be thicker for larger ice-grains
and this is to be expected since solid-state ices
composed of larger ice-grains are characterized by larger viscosities.

We also note there there are critical values of the ice-grain
size for which the transition into convection occurs
at depths for which the
base temperature, $T_b \equiv T_s + H \overline T_{_{0z}}$, 
corresponds to the melt temperature $T_m$ at
the triple-point for \N2.
The bottom panel of Figure \ref{Viscosity_contrast_for_Convection}
shows the results of a model in which the surface
temperature $T_s = 38.5$\Kelvins \ and
where $\overline T_{_{0z}} = 20$\Kelvins/km, 
and we see that $T_m$ corresponds to about 63 \Kelvins \
for ice-grains of size $d_g = 2-3$ mm (depending upon
which boundary condition is employed).  Cross-referencing
this critical grain-size with the corresponding critical
value of the depth of the layer upon which transition
is predicted to occur
(Fig. \ref{Min_Layer_thickness_for_Convection})
shows that this happens at layer thicknesses $H\approx 1.2$ km.

\par
\bigskip
We conclude by noting that the onset of convection in sheared flows is
known to occur at (and often significantly) higher values of Ra as compared 
to their non-sheared
equivalents, e.g., see the discussion and references in 
\cite{Landuyt_Ierley_2012,Goluskin_Spiegel_etal_2014}.  
As such, we consider
the Ra numbers for transition to be lower bounds for the onset of convection
in down-gradient flowing \N2/CO ice.  In practice this means for us that
flowing ice layers probably move laminarly down-gradient for layer thicknesses
greater than the limits suggested by conditions indicated in Fig.
\ref{Min_Layer_thickness_for_Convection}.  When revisions to \N2 rheologies become
available, many of the figures quoted in this work will need to be revised.


\begin{figure}
\begin{center}
\leavevmode
\includegraphics[width=8.85cm]{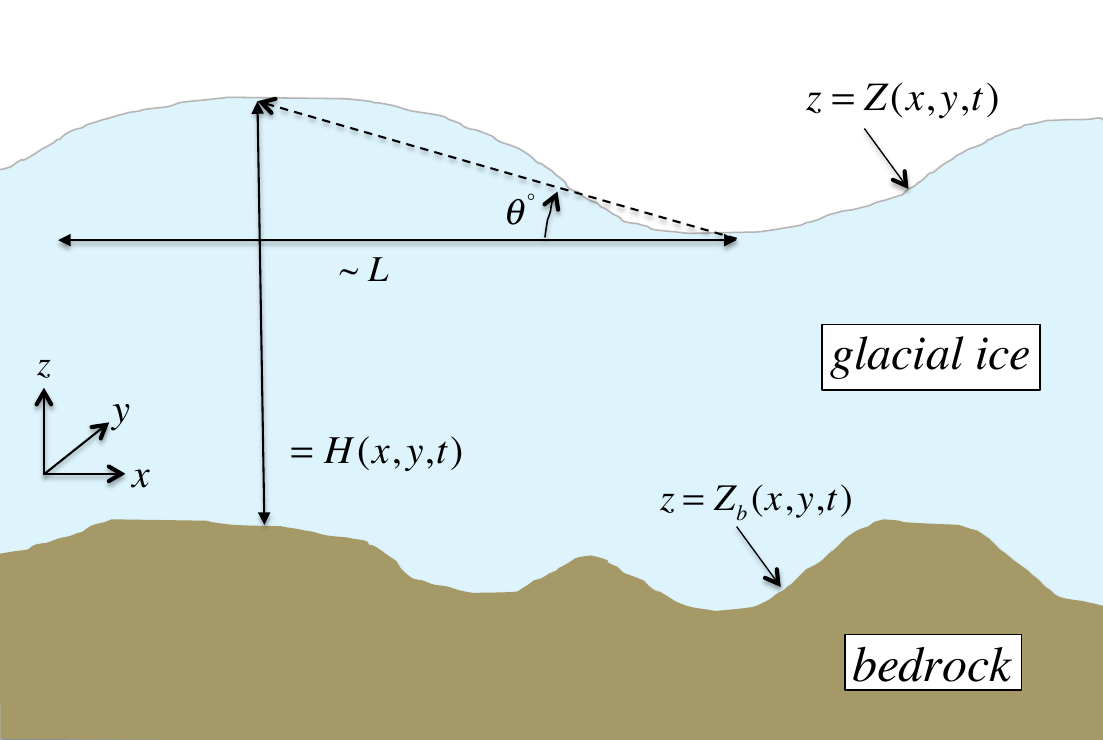}
\end{center}
\caption{Glacial ice modeling schematic.  Evolving glacial ice of thickness $H(x,y,t)$
with surface elevation position $z=Z(x,y,t)$ and bedrock surface $z=Z_b(x,y,t)$.  All quantities
are space and time dependent.    The angle $\theta$ describes the grade of the surface
feature with horizontal scale $\sim L$.
}
\label{Modeling_Schematic}
\end{figure}

\section{Landform evolution modeling}\label{glacial_flow_modeling}

Partly motivated by our findings regarding the onset
of convection described in Section \ref{convective_onset},
we treat the glacially flowing ice as a single laminarly flowing viscous fluid in the infinite
Prandtl number limit. 
%
This assumption permits the use of
vertically integrated modeling methods (e.g., see
discussions in \cite{Hindmarsh_2004,Benn_Evans_2010}) whose
asymptotic validity is rooted in perturbation analyses (e.g., \cite{Balmforth_etal_2003}).
The assumption of laminar flow is valid so long as the moving glacial ice layers
are not thick enough to experience buoyant convection.   Vertically integrated
modeling is expected to be valid so long as the horizontal length scale of the layer dwarves the vertical extent of the layer -- in the glaciology literature 
this is often referred to as
the shallow sheet approximation (SSA) \cite{Hindmarsh_2004}.
\ref{glacial_flow_model_development} contains
a detailed exposition of the derivation of the vertically integrated SSA model we use and
interested readers are directed there.

The model setup is sketched in the cartoon found
in Fig. \ref{Modeling_Schematic}.  We consider a single spatiotemporally
varying ice component whose local vertical thickness as a function of horizontal
location is given by $H(x,y,t)$.  The ice sits on a bedrock
whose surface is located at $z=Z_b(x,y,t)$.  Note in \ref{glacial_flow_model_development}
we develop a model in which the bedrock may be convertible into ice due
to various landform modifying processes, but for our purposes here in this
study we treat the bedrock as static.  The local elevation is given
by $z=Z(x,y,t) = H(x,y,t) + Z_b(x,y)$.  The evolution of the local ice
layer is given by the (scaled) horizontal divergence of the corresponding
mass flux
\beq
\partial_t H = \nabla\cdot {\bf q},
\label{fundie_equation_of_paper}
\eeq
where the {\emph{total mass-flux}} is given as the sum
of a ``dry" and ``basally wet" glacial flow,
 \beq
 {\bf q} = q\sub 0 \nabla Z, \qquad
 q\sub 0 = q\sub {sl} + q\sub{{\rm dry}}.
 \label{bold_q_def}
 \eeq
 The dry component of the mass-flux coefficient is given by
 \beqa
 & & q\sub{{\rm dry}} = 
 g\sub{{\cal Q}}\exp\left[\frac{ H/H_a}{1 + H/H\sub{\Delta T}}\right]q\sub{{\rm glen}} ,
 \nonumber \\
& & q\sub{{\rm glen}} \equiv 
\frac{A_s \big(\rho_s g H\big)^n H^2}{n+2}
 {S^{n-1}}{}.
 \label{qdry_appx_def}
 \eeqa
 The model adopts a non-Newtonian Glen law formulation (Section 2.2) modified
 to take into account the strongly insulating nature
 of nitrogen ice (Section 2.1). The classical Glen law part of the expression
 is contained in $q\sub{{\rm glen}}$ where the Glen law index is $n$.  $S$ is the local scalar tangent
 of the surface, i.e., $S \equiv |\nabla Z|$.
 $A_s = A(T_s)$ is the value of the Arrhenius prefactor
 expression, found in equation (\ref{Glen_Prefactor_N2}),
 evaluated at the surface temperature $T_s$.  
 The parameters appearing in the dry glacial
mass flux formula, equation (\ref{qslide_def}), $H_a$ and $H\sub{\Delta T}$, are given by
\[
H\sub{\Delta T} = T_s/\overline T\sub{0z},
\qquad 
H\sub a = T_s^2/(T\sub a \overline T\sub{0z}),
\]
where $T\sub a, \overline T\sub{0z}$ are respectively,
the activation temperature of \N2 ice (Section 2.1) and
the vertical temperature gradient inside the ice (Section 2.3).
The order 1 numerical parameter $g\sub{{\cal Q}}$, which is a corrective term
depending upon the rheology of the flowing material, is
described in detail in \ref{glacial_flow_model_development}. For
all of our subsequent modeling we
 adopt a constant value of $g\sub{{\cal Q}} = 0.5$.

 The mass flux coefficient of the basal sliding component is
 given by
 \beq
 q\sub{sl} 
=
 \frac{H}{|\nabla Z|} \left\{
\begin{array}{cl}
 u\sub{{\rm sl}0}, & H > H_m, \\
{\bf 0}, & 0< H  < H_m.
\end{array}
\right.
\label{qslide_def}
 \eeq
The physical meaning of the model is that once a minimum
thickness $H_m$ is achieved, the base of the layer experiences
melt.  $H_m$ generally depends upon many physical inputs including latitude, local sloping angle of
the surface and the emergent geothermal flux (i.e., $\overline T_{0z}$), 
but for our purposes in this study we adopt a constant value for it.  
The wet base adds a constant velocity
 $u\sub{{\rm sl}0} {\bf {\hat t}}$
to the flowing glacier, where ${\bf {\hat t}} \equiv \nabla Z/|\nabla Z|$
is a unit vector indicating the map-projected
direction of the local downslope of the ice.   The speed $u\sub{{\rm sl}0}$ is
the magnitude of the basal flow which we treat as a model input parameter,
but it may generally be considered to be a function of other inputs
like total overburden pressure and temperature.  This physical model
is derived from terrestrial glacial flow modeling\footnote{ Caution is advised. It is
likely that a basally wet \N2 ice layer may behave
differently than basally wet terrestrial \water ice glaciers owing primarily to the lower density of liquid \N2
compared to its solid phase.  Sufficiently thick layers of low viscosity liquid \N2 underneath
an ice layer should
lead to Rayleigh-Taylor instabilities 
which would, consequently, lead to vertical mixing of the liquid \N2
with the overlaying ice.  Further theory is required in this instance.}.

We reflect upon the functional dependence
 on the dry mass-flux coefficient $q\sub{{\rm dry}}$, which scales as 
 \[
 \sim H^{n+2} \exp\left( \frac{H/H_a}{1+H/H\sub{\Delta T}}\right),
 \]
indicating its significant deviation from the classical Glen Law form
 used for terrestrial glacial ice modeling, i.e., the classical
 form contained in the expression $q\sub{{\rm glen}}$.  We refer
 to this composite form as the Arrhenius-Glen mass-flux dependency. 
 This new form reflects the Arrhenius
 prefactor of \N2's rheology, which becomes significant 
 because of the strongly insulating
 nature of \N2 that promotes the development of strong temperature gradients 
 inside a conducting layer of \N2 ice.  This occurs on a relatively short period of time
 (see discussion in Section \ref{thermal_properties}).
 \par
 \bigskip

 \begin{figure}
\begin{center}
\leavevmode
\includegraphics[width=9.0cm]{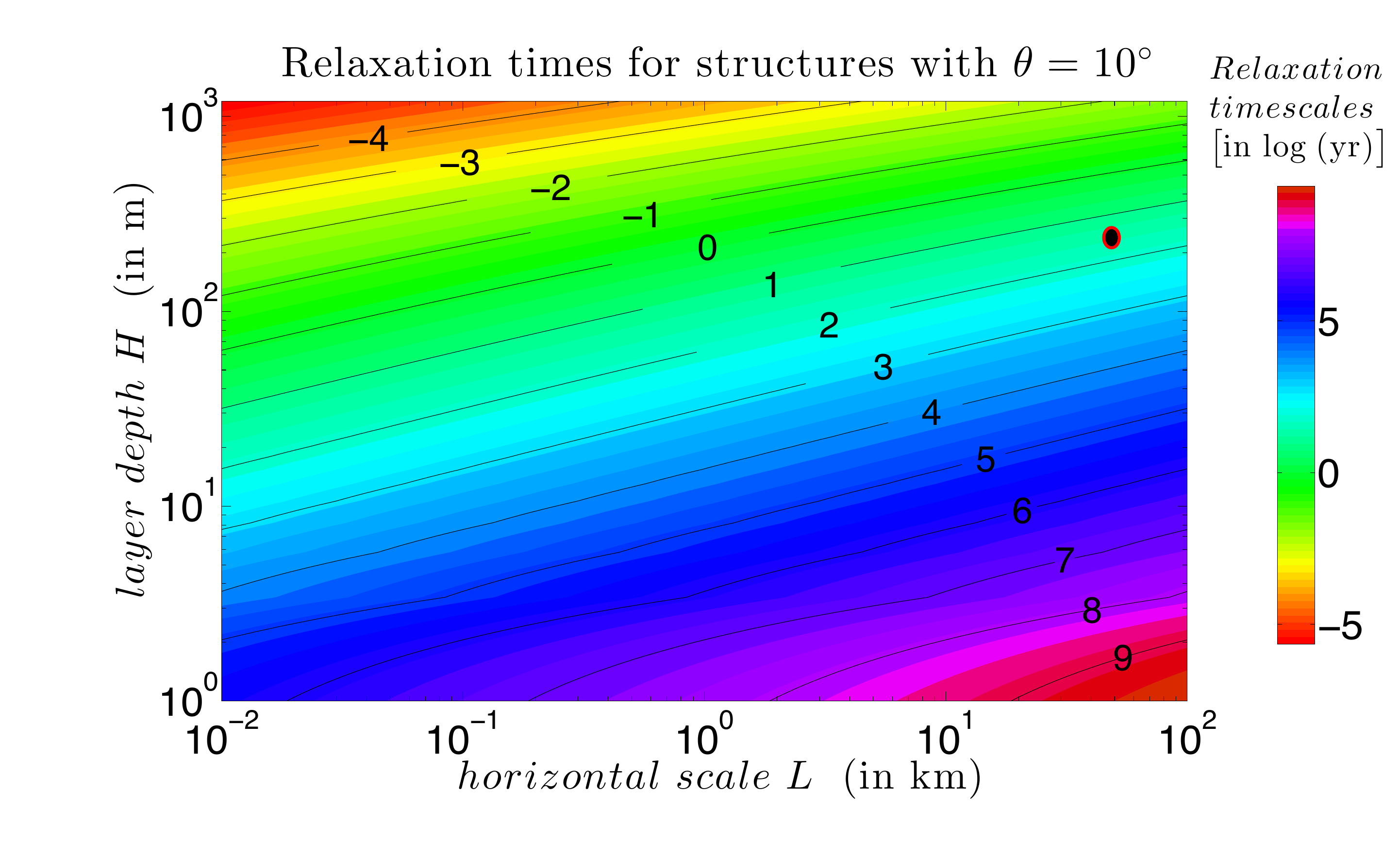}
\end{center}
\caption{Flow timescales $\tau$ for glacial ice structures
composed of \N2 ice characterized by the rheology of \cite{Yamashita_etal_2010}:
glacial ice configurations with surface
sloping grades of $\theta = 10^\circ$.  Flow timescales are presented for
a range of horizontal and layer-depths $L$ and $H$ respectively.  
The surface temperature of the ice is $T_s = 38.5$\Kelvins \ 
with a vertical temperature gradient $\overline T\sub{0z} = 20$\Kelvins/km. 
Other parameters are $n=2.1$, $H\sub a \approx 175$m, 
$H\sub{\Delta T} = 1925$m with $g\sub{{\cal Q}} = 0.5$.}
\label{Relaxation_times_1}
\end{figure}

\section{Some model results}
We present here a preliminary examination of the two questions/scenarios
posed in the Introduction.  We work with model landscapes adapted
from digital elevation models (DEM) of two regions on the eastern side of
SP (further detailed below in section \ref{dem_model}).  We treat these surfaces as the primary bedrock 
and onto them
we add a certain amount of \N2 ice and observe the response.
We solve eqs. (\ref{fundie_equation_of_paper}-\ref{qslide_def})
using a stripped down version of MARSSIM \cite{Howard_2007} in which
the glacial flow law is assessed using both either a second
or fourth order correct finite-difference flux method.
The program solves the equations of motion in a doubly periodic domain.

\subsection{Flow timescales - some estimates}\label{glacial_flow_rate_timescales} 
\par
We can make certain estimates regarding the flow time-scales
associated with various ice-configurations assuming dry \N2 ice layers.
We adopt the rheology of \cite{Yamashita_etal_2010} -- but we keep in mind
that all estimates made here will need to be revised when future improvements
to \N2 ice's rheology are made available.
\par
\begin{figure*}
\begin{center}
\leavevmode
\includegraphics[width=9.0cm]{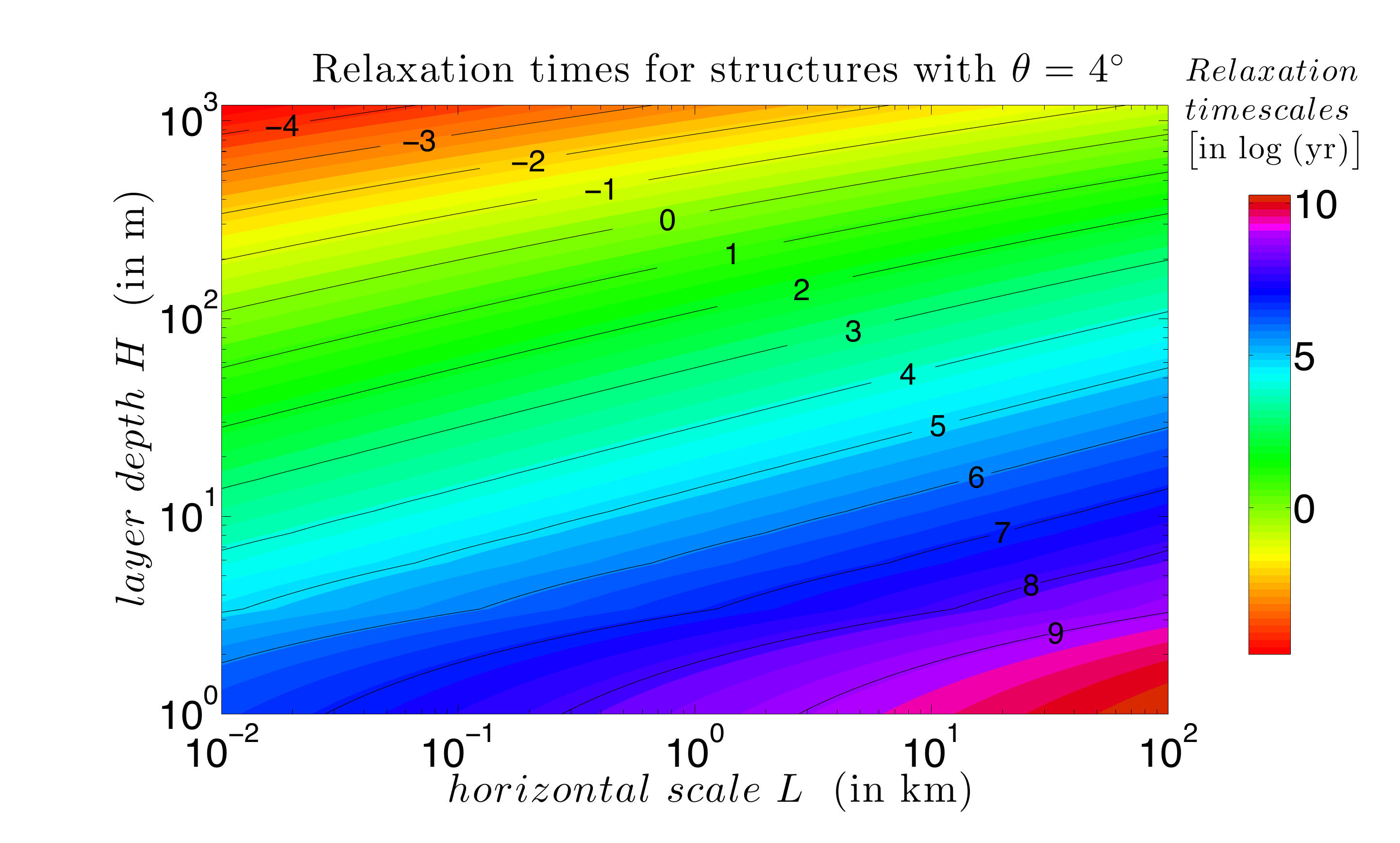}
\includegraphics[width=9.0cm]{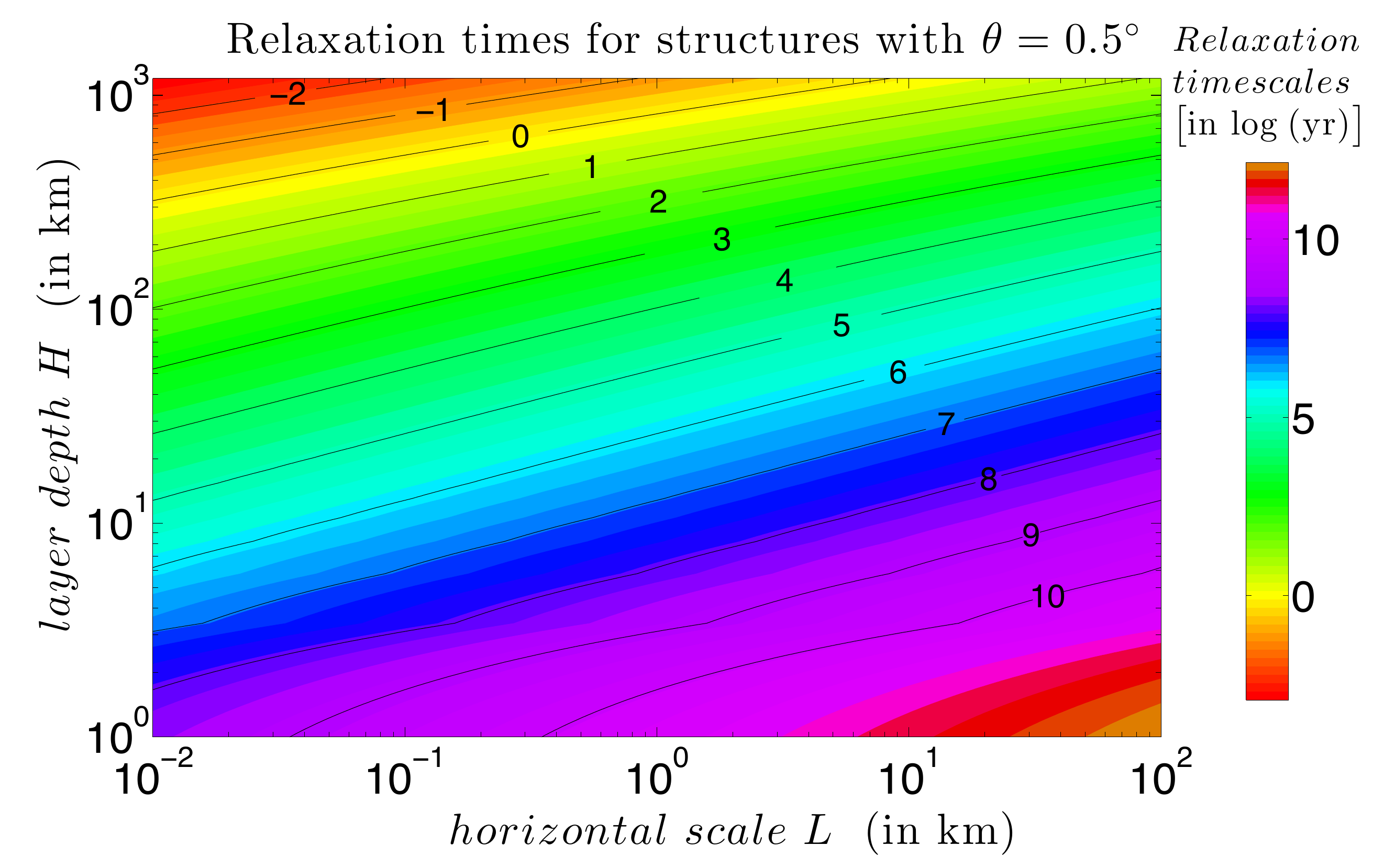}
\end{center}
\caption{Like Figure \ref{Relaxation_times_1} except for ice configurations with surface
sloping grades of $\theta = 4^\circ$ (top panel) and
$\theta = 0.5^\circ$ (bottom panel).}
\label{Relaxation_times_2}
\end{figure*}

An e-folding flow timescale, $\tau$, may be assigned to surface features of horizontal length $L$,
vertical height $H$, and
surface gradients characterized by the grade angle $\theta$.
Assuming that the bedrock is static,
temporal variations in $Z$ are the same as temporal variations in $H$.  Thus, the
left hand side of the evolution equation (\ref{fundie_equation_of_paper}) may be approximated
by $H/\tau$.
 Using the form for the mass-flux given by Eq. (\ref{bold_q_def})
together with the approximate form for the dry \N2 ice mass-flux $q\sub{{\rm dry}}$
in eq. (\ref{qdry_appx_def}), and the assumption
that horizontal variations of surface features scale with $L$ (see Fig. \ref{Modeling_Schematic}), 
the order-of-magintude measure of the right-hand-side
of Eq. (\ref{fundie_equation_of_paper}) is given by $\sim L^{-1}|q_0\tan\theta|$.
Putting these two estimates together produces the following expression for the rate
 $\tau^{-1}$,
\beqa
 {\tau^{-1}} 
\approx 
\frac{|q_0 \tan\theta|}{HL}, \hskip 5.0cm  & & \nonumber \\
 =
g\sub{{\cal Q}}\frac{H}{L}\frac{A_s \big(\rho_s g H\big)^n}{n+2}
 \exp\left[\frac{ H/H_a}{1 + H/H\sub{\Delta T}}\right]
 \big|\tan\theta\big|^n . & &
\eeqa
We note that as written, $L/\tau$ approximates the surface velocity of the flowing
glacier ice.
In Figures \ref{Relaxation_times_1}-\ref{Relaxation_times_2} we present color contour
plots for the relaxation times associated with a variety of glacial structures
with various values of $H$, $L$ and $\theta$ (see Fig. \ref{Modeling_Schematic}).
\par
Some noteworthy figures emerge.
The black oval
in Figure \ref{Relaxation_times_1}
 designates the e-folding relaxation time associated with a 50km long
channel, sloping at $\theta = 10^\circ$ and initiated with 200 m of glacial ice.
The corresponding relaxation time is about 50 years, similar
to the figure reported in \cite{Moore_etal_2016}.  One may interpret
this time-frame equivalently as the e-folding period on which
such a channel drains out a $(1-1/e)$ fraction of its initial ice content.
Imagining that we indeed begin with 200 m \N2 thick layer and after
the just described e-folding timescale, we would have remaining an ice layer of about 75 m thick
which, according to 
Figure \ref{Relaxation_times_1},
ought to undergo another $(1-1/e)$ fractional
drain in about 5000 years.  Of course, this difference in drainage times is
due to the strong nonlinear dependence of the mass-flux on $H$, namely
the Arrhenius functional dependence shown in Eq. (\ref{qdry_appx_def}).  Nonetheless,
and all other things being equal (especially the surface temperature $T_s$),
any substantial amounts of ice 
accumulated in such drainage channels are 
likely to drain out to levels (say, below 50 meters) 
in well under 100-500 years ($\sim$ 0.5 - 2 Plutonian orbits).
Similar timescale estimates can be made for structures and channels of differing
extents and depths.

\begin{figure*}
\begin{center}
\leavevmode
\includegraphics[width=17.9cm]{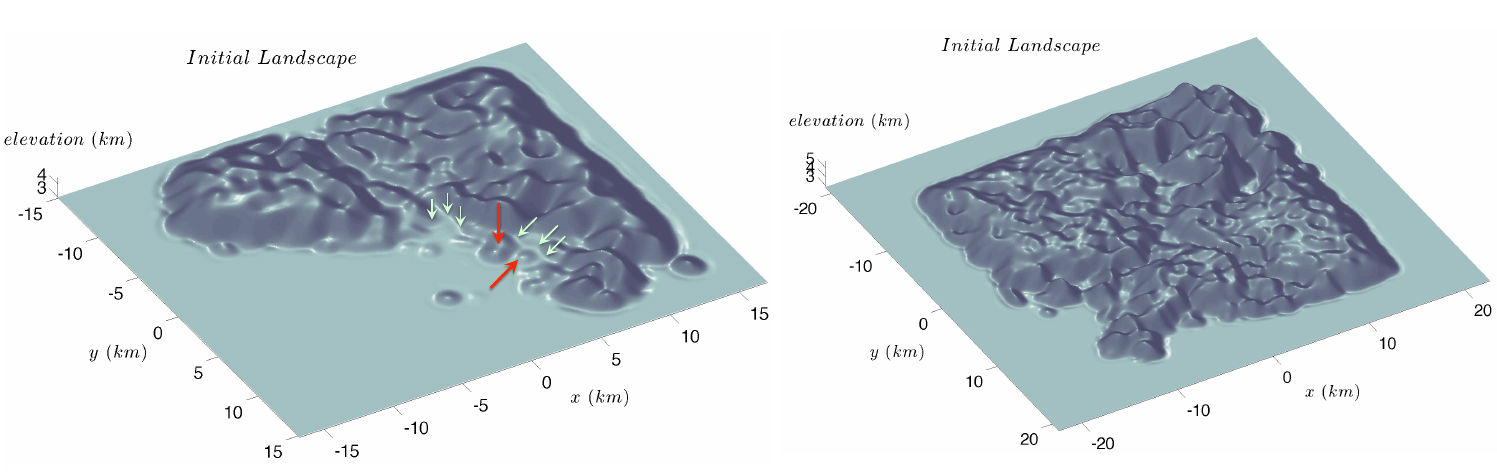}
\end{center}
\caption{Digital elevation models of two eastern sections of SP called GL-1
(left panel) and GL-2 (right panel).
 The topography
has been artificially flattened along borders of frame to facilitate
modeling. This surface is treated as the bedrock upon which we examine
various glacial flow responses.
 Reference topography is indicated for GL-1: local highs (red arrows)
 and lows (lime arrows).  This depiction of the
 topography simulates lighting coming from the top left of both images.}
\label{GL-12_InitialBedrock}
\end{figure*}

\par
\begin{figure*}
\begin{center}
\leavevmode
\includegraphics[width=0.99\textwidth]{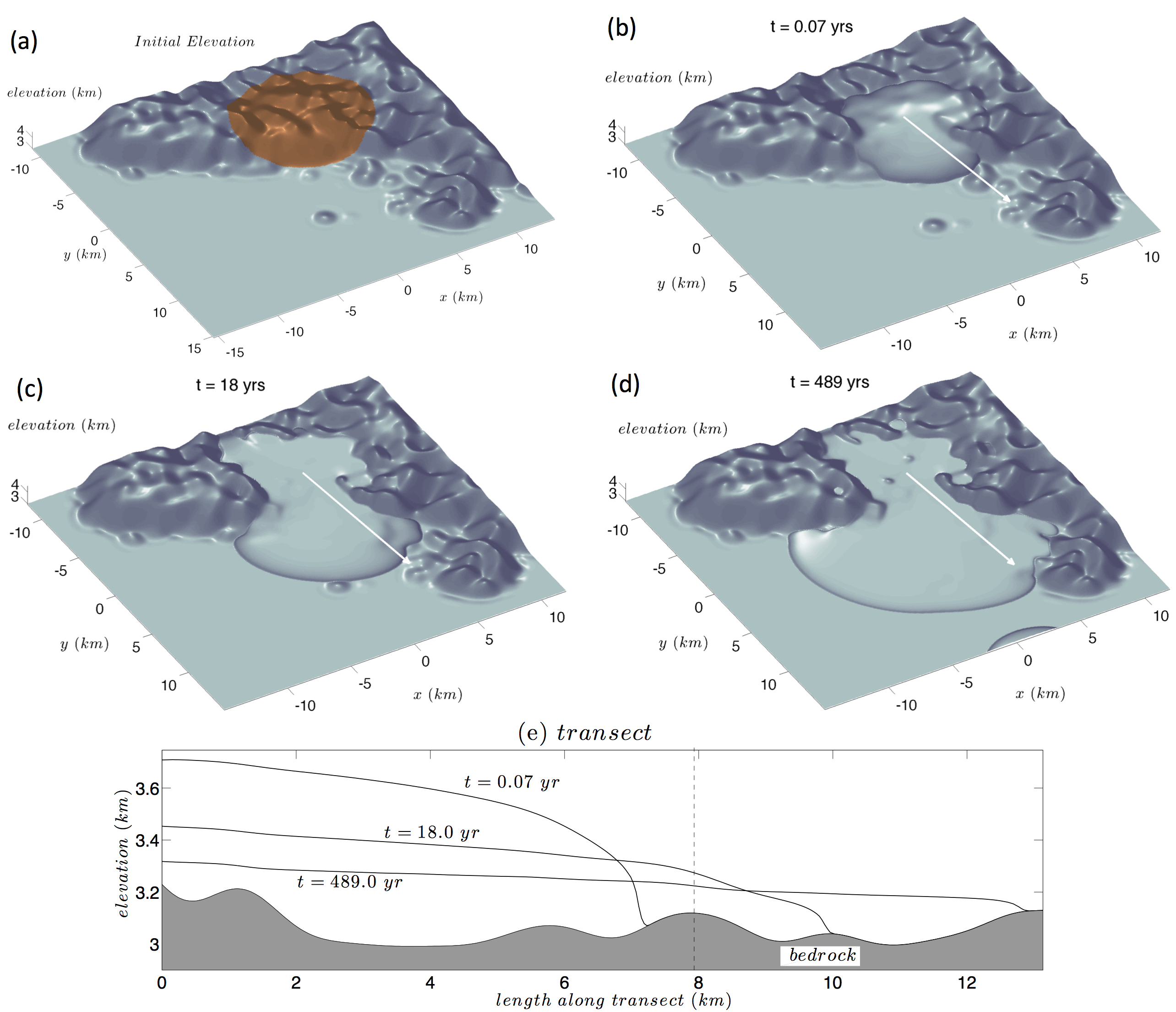}
\end{center}
\caption{Glacial flow evolution of dry \N2 ice over landscape GL-1 with
power law coefficient $n=2.1$ and initial mound amplitude of $H\sub{r0} = 0.7$km.  Panel (a) shows initial \N2 ice distribution 
designated by orange highlight.  Panel (b) shows early readjustment of initial configuration ($t = 0.07$ yrs).
Panel (c) and (d) show the late time development of the elevation as glacial flows over topography.  
Flow over topography is especially pronounced over the ridgeline indicated
by the red arrow in left panel of Fig. \ref{GL-12_InitialBedrock} and at transect position $\sim 8.0$km in panel (e).   
By late times ($t=489$ yrs), the drop in elevation diminishes and the imprint
due to the ridgeline topography becomes increasingly imperceptible.}
\label{GL_1_LeakyHeights}
\end{figure*}

\subsection{Sample initial surfaces from digital elevation models}\label{dem_model}
As we are interested in the character of \N2 ice flow on Pluto, we have
adopted two regions along the eastern shoreline of SP
to use as the topography on which the glacial ice flows.
These two rectangular regions
called GL-1 and GL-2, the locations with respect to SP of which are indicated
in Figure \ref{LandScapeLocator}, 
are extracted from DEMs created using stereogrammetry -- a method of elevation determination
applied successfully in many previous studies focusing on icy satellites
\cite{Schenk_2002,White_etal_2013}.
The master stereo DEM was created from 
from 0.495 and 0.32 km/px observations
\footnote{Specifically, these are the P$\underline{\ \ }$MPAN1 observation (resolution
0.495 km/px) and the P$\underline{\,\ \ }$MVIC$\underline{\ \ }$LORRI$\underline{\ \ }$CA observation (resolution 0.32 km/px).}
--
both of which are observations obtained using the MVIC camera.  
The pixel scale we adopt therefore is 0.495 km/px.  The precision in relief
for these DEMs is about 225 meters \cite{Moore_etal_2016}.

\par
The initial bedrock landscapes used in our simulations, shown in 
Figure \ref{GL-12_InitialBedrock},
are primed for use based on the following procedure.
Because the stripped-down version 
of MARSSIM we use solves the equations of motion in a doubly periodic domain,
the landscape should have equal elevations and gradients
along the borders of the computational domain.  We therefore take the raw DEM, $Z\sub{{\rm raw}}(x,y)$ and apply
a filter which forces the elevation to be equal along the borders of the rectangular section as
well as forcing the gradients there to be zero.  We simultaneously impose a minimum
elevation requirement at some arbitrary level $Z_{min}$, i.e.
\[
Z\sub{{\rm raw}} \rightarrow Z\sub{{\rm b}} = {\rm max}\big[Z\sub{{\rm raw}},Z\sub{{min}}\big].
\]
We always choose the elevation on the computational border to be equal to the minimum elevation $Z_{min}$.
 The resulting landscape looks like an island as can be seen in the figures.  
The choice of $Z_{min}$ is guided by our desire to have some interesting topographic
features appearing along the shoreline of these island-like landscapes
(like, for instance, the topography indicated
by red arrows on the top panel of Fig. \ref{GL-12_InitialBedrock}).
The final stage involves doubling the resolution of the model landscape
using a third order spline method.  This is done in order to smooth out
small (pixel to pixel) scale scatter 
that is a byproduct of most DEM methods including stereogrammetry.
\cite{Schenk_2002,White_etal_2013}.
The grid resolution of each of the two landscapes GL-1 and GL-2 are respectively,
$N_x\times N_y = 304\times 340, \    424\times 452 $.

\subsection{Some numerical results}
We discuss here three suites of solutions.  Each landform evolution simulation
begins with a mound of \N2 ice centered on at the map position $(x,y) = (x_c,y_c)$,
with a form given by
\beq
H\sub0 = H\sub{r0} \exp\left[\left(\frac{r}{\delta r}\right)^m\right],
\eeq
where
$r = \sqrt{(x-x_c)^2 + (y-y_c)^2}$.  All initial profiles use a power index $m=6$.  
The resulting profile looks like an inverted dinner plate
where ice regions close to $r=0$ are nearly flat with thickness $H\sub{r0}$, while a precipitous drop off occurs
at the value of $r$ approaching the nominal mound size, $\delta r$.
All numerical simulations adopt the non-Newtonian Arrhenius ice rheology of annealed \N2 
\cite{Yamashita_etal_2010}.  Thus, for $q\sub{dry}$,
we use the formula
eq. (\ref{qdry_appx_def}), 
wherein we adopt the ice surface temperatures to be $T_s = 38.5$\Kelvins,
together with $\overline T\sub{0z} = 20$\Kelvins/km and 
$T_a = 425$\Kelvins -- which taken together leads to
$H\sub a = 0.175$km and
$H\sub{\Delta T} = 1.95$km.

\begin{figure*}
\begin{center}
\leavevmode
\includegraphics[width=0.99\textwidth]{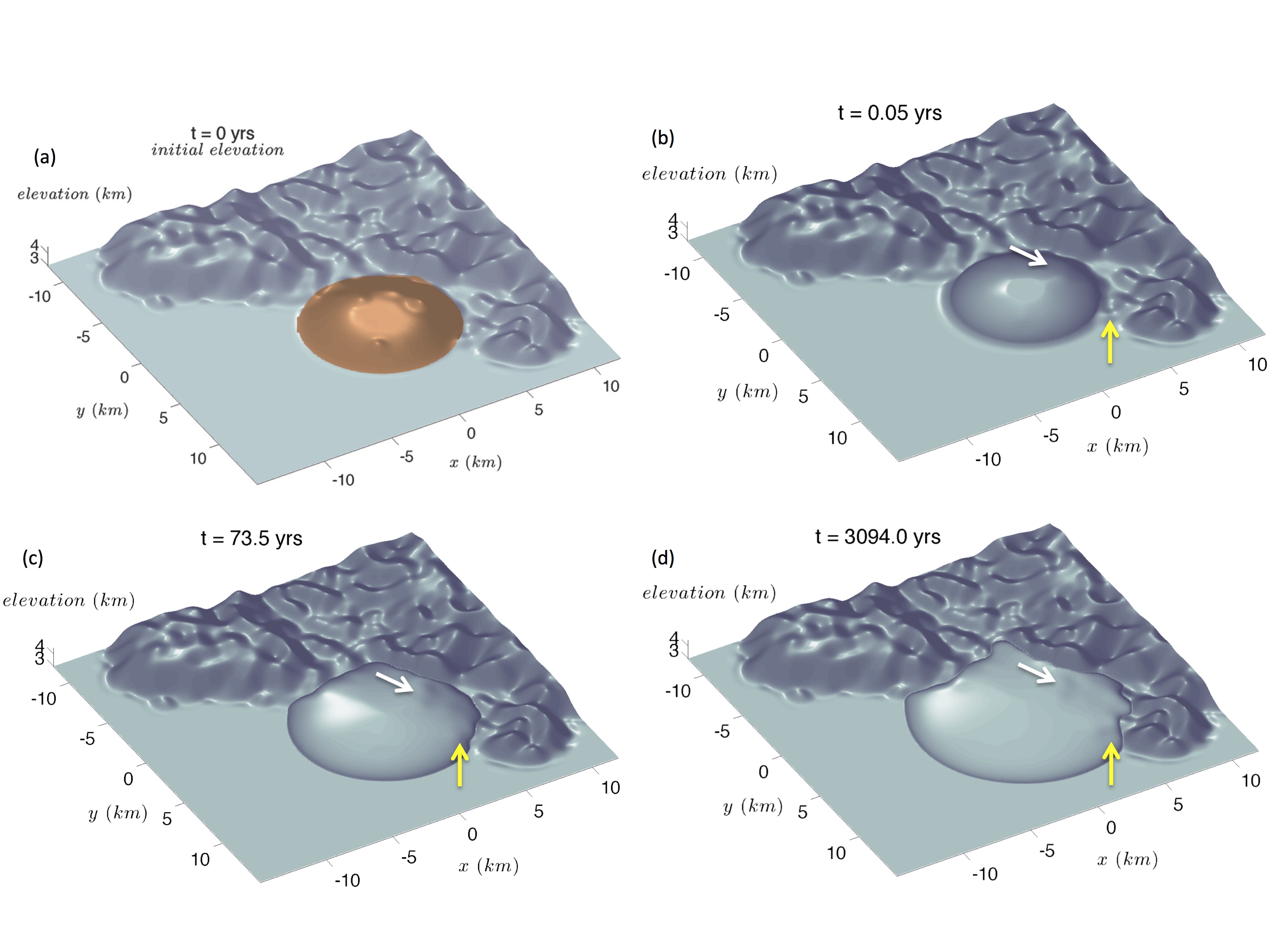}
\end{center}
\caption{Glacial flow evolution of dry \N2 ice toward landscape GL-1.
Flow parameters are
$n=2.1$ and $H\sub{\Delta T} = 1.95$km and $H_a = 0.175$km
with
an initial mound amplitude of $H\sub{r0} = 0.7$km.
  Panel (a):
Initial ice mound
emplaced on flat part of bedrock (indicated by orange) is allowed to relax. Panel (b-d): 
early and late time configurations of this relaxation process.  Topographical imprint
upon the flowing glacier is apparent (white and  yellow arrows).  The imprint is relatively
weak for this rheology.}
\label{GL_1_Exp2_REVISED}
\end{figure*}


 Figure \ref{GL_1_LeakyHeights} displays the results of a numerical simulation
 in which the initial \N2 ice mound with $\delta r = 4.3$km is emplaced in the highlands of landscape
 GL-1 ($x_c = 3\ $km, $y_c = -\ $4km), with a very large ice thickness $H\sub{r0} = 0.7\ $km.
 Despite the large value of  $H\sub{r0}$, the layer is evolved as a dry glacier 
 (i.e. no basal melting)
 because
 the purpose
 of this demonstration  is to exhibit how flowing glacial ice reflects bottom topography.
 After an initial adjustment phase of this advancing dry glacier ($<1$ yr),
 \footnote{All numerical simulations performed here have been initiated
 with relatively ``unnatural" initial ice distributions, in the sense that they represent ice accumulations
 that are not those realizable from some quasi-steady process.  As such, in all such cases, the simulations
 go through two typical stages: (i) a rapid initial readjustment phase followed by
 a (ii) slower creeping relaxation stage.  In order to avoid
 numerical instabilities, this initial rapid
 readjustment phase must be evolved with relatively small time steps.  After a sufficient
 amount of time in which the slowly creeping relaxation phase
 is reached, the time-stepping of the simulations may then be increased -- generally by a factor of 5-10 times
 the initial time step.} 
 significant ice drainage from the uplands
 and onto the flats occurs in under 10-20 years --
 both timescales which are consistent with the glacial timescale calculation in section
 \ref{glacial_flow_rate_timescales} as well as the numbers quoted in \cite{Moore_etal_2016}.
 \par
 There are several notable features in the results. 
 Most prominently is how the ridge-line indicated
 by the large arrow in Fig.\ref{GL-12_InitialBedrock} is clearly imprinted on the surface ice
 once it has covered the ridgeline.  The dark elongate feature in Fig. \ref{GL_1_LeakyHeights}c,
 that is cut by the white transect arrow, is the expression of the aforementioned ridgeline.
 This elongated feature on the surface ice indicates a drop in the ice level of about 10-20 meters
 as can be seen in Fig. \ref{GL_1_LeakyHeights}e.  {  Physically speaking, this drop in the
 ice level at the 8km mark along this transect indicates the effect of
 ice piling up as it moves over and around the subsurface ridge.}
 There are other similar elongated dark features that also similarly
 reflect short lengthscale changes in the bottom topography (yellow arrow).
 By late times, $t=489$ yrs Fig. \ref{GL_1_LeakyHeights}d,
 this surface feature weakens because the drop in the ice level
 reduces the imprint to nearly imperceptible amounts.  
 \par
 These trends suggest
 that the wavy elongated features seen on the northern shores of SP, especially those seen in Fig. \ref{Outflow_Image_1}b,
 are not only imprints of bottom topography but are ephemeral as well. 
 We note here that the imprint is relatively diffuse and this is likely because of the
 more fluid-like quality of this material due to its low value of the power law index $n$ (see below).
Nevertheless, if the wavy darkened features seen near the northern shoreline of SP 
are indeed imprints of topography, then this zone
may have experienced a recent surge of glacial activity (in the last few hundred years).
Despite the more fluid-like nature of the ice, the advancing glacial front, descending
from the pitted highlands and passing through the basin and out onto the flats of the plain,
is reminiscent of terrestrial glaciers that have power law indices closer to the range
of $n=4-6$.

\par

\begin{figure}
\begin{center}
\leavevmode
\includegraphics[width=0.48\textwidth]{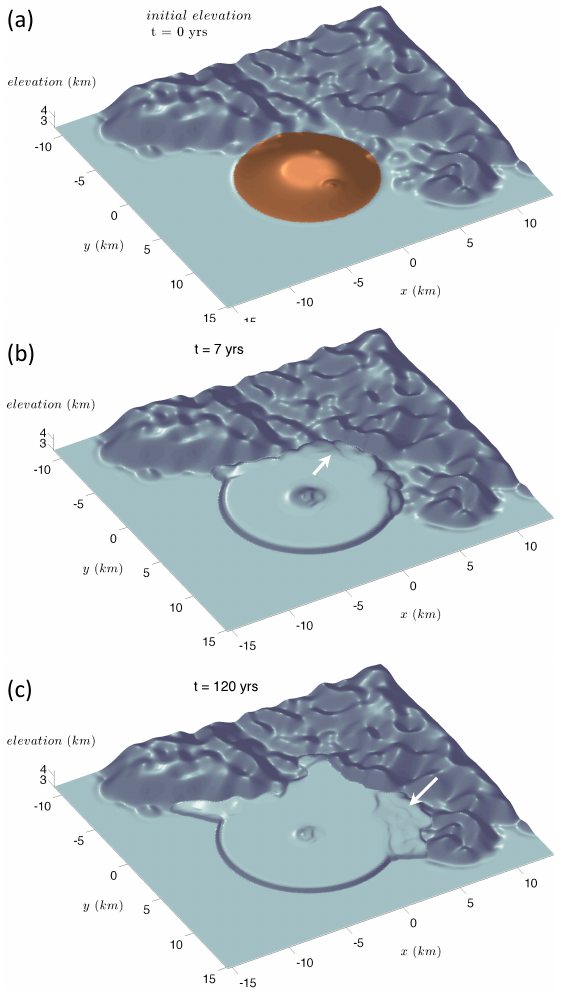}
\end{center}
\caption{Glacial flow evolution of dry \N2 ice toward landscape GL-1.
Similar numerical experiment to Figure \ref{GL_1_Exp2_REVISED} except for a stiffer
rheology (see text). Panel (b-c): 
early and late time configurations of this relaxation process.  Note the appearance of wavy
patterns (white arrows) and their qualitative resemblance to similar undulating
patterns seen near the northern shore of SP (Fig. \ref{Outflow_Image_1}).  Wavy
patterning is an imprint of local topographic extremes shown by the green arrows
in the left panel of Fig. \ref{GL-12_InitialBedrock}.  The wavy pattern indicated
by arrow in panel (b)
is no longer visible at later times, panel (c).}
\label{GL_1_Exp2_QuasiBingham}
\end{figure}

\begin{figure*}
\begin{center}
\leavevmode
\includegraphics[width=0.99\textwidth]{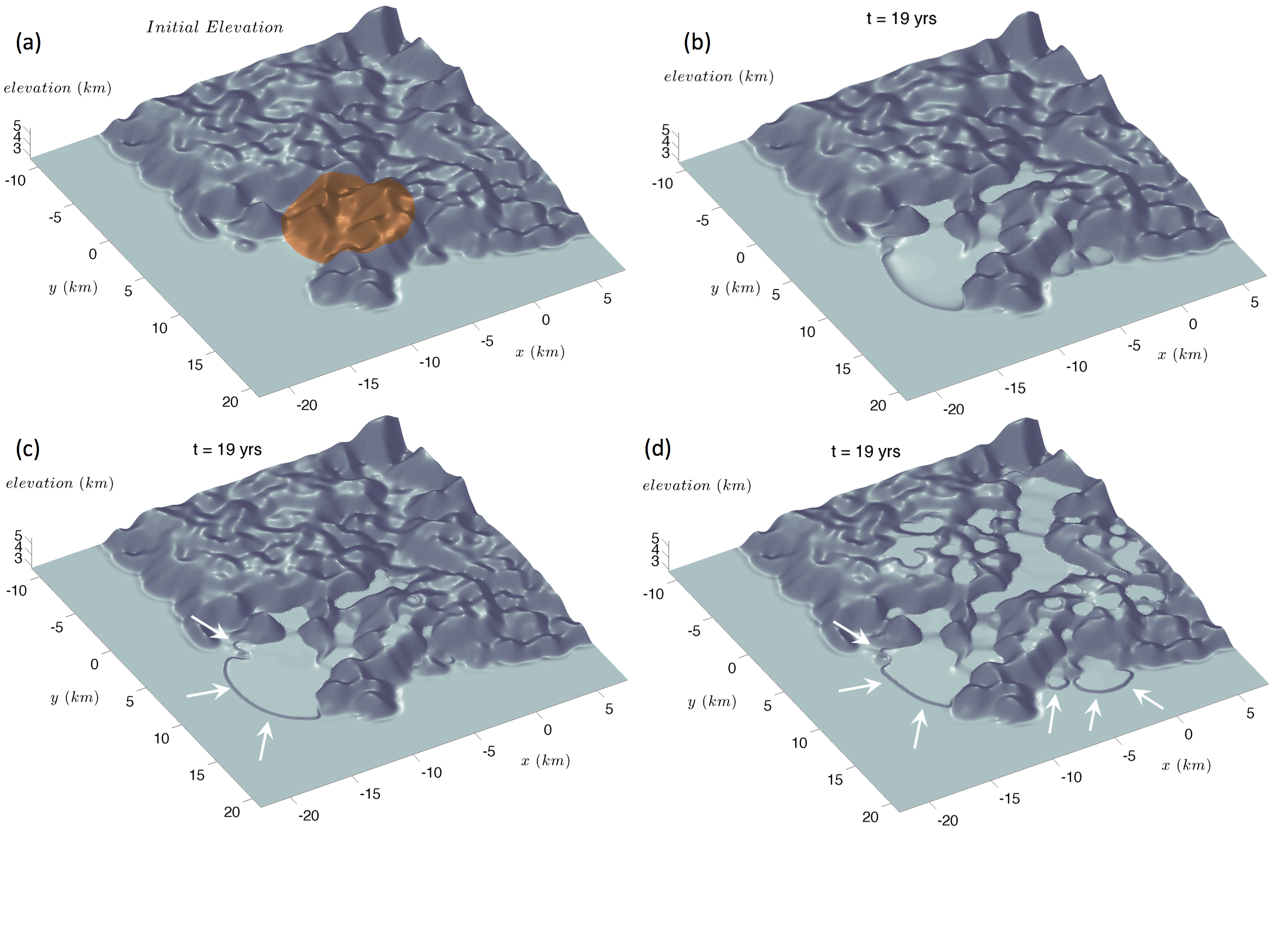}
\end{center}
\caption{Glacial flow evolution over landscape GL-2 of either basally wet or dry \N2 ice
with $n=2.1$ and $H\sub{\Delta T} = 1.95$km and $H_a = 0.175$km
and initial mound amplitude of $H\sub{r0} = 0.21$km.
Panel (a) exhibits the initial ice mound emplaced on sloping highlands which
is then allowed to relax.
The resulting shape after 19 years is shown for the case of no
basal slip, panel (b); and with basal slip, panel (c).  
In the latter - slip occurs with a speed $u\sub{sl0} = 0.2$km/yr once the
the local ice layer thickness exceeds $H>H_m = 100$m.  Panel (d) depicts
a similar numerical experiment involving basally wet glacial flow, 
but for a wider shallower initial ice mound (details
in text).  Owing to the relatively low power law index, simulations of both of basally wet 
and dry glacial flow develop classical flow lobe features
once ice has reached the flats.  Flow lobes are more pronounced
in basally wet flow simulations.   The numerical experiment in panel (d) shows how
glacial ice collects inside local topographic lows as well.
}
\label{GL_2_Composite}
\end{figure*}

\medskip
Figure \ref{GL_1_Exp2_REVISED} displays results of a similar experiment like shown in the previous
figure except the original mound is placed on the flat region
of landscape GL-1 ($x_c = -2\ $km, $y_c = \ $4km) --  all other parameters and
conditions are otherwise the same. In this instance the flow moves toward
the shoreline, yet we see similar qualitative features of topographic imprinting
on the moving ice as it begins to envelope exposed relief.  Interestingly, the patterning
takes on a wavy character once flow begins to move into the mouth of the basin
around $x=0$ km, $y=0$ km.  The wavy feature reflects the local topographic
low of the hills in the mouth of the basin -- as pointed out by the green arrows 
in the left panel of Fig. \ref{GL-12_InitialBedrock}.  As before, both the wavy features
eventually disappear as the \N2 ice flattens out over time
and the relief of the topography imprinted upon the ice is fairly muted.
\par
 By contrast, in Figure \ref{GL_1_Exp2_QuasiBingham} we show the result
 of a similar setup but with a relatively stiff rheology based on a quadi-Bingham
 flow model used in describing the features on the surface of Helene
 \cite{Umurhan_etal_2015}.
 The purpose of this particular demonstration is to show that
 the wavy features indicating bottom topography are brought out in 
 starker relief for surface flow models with stiffer rheology.  The long time
 fate is the same: the wavy features eventually fade over time as the 
 surface ice relaxes.
 The actual properties of \N2 ice rheology, possibly contaminated by other
 pollutants like \CH4 and tholins, remain to be constrained.
\par
We also note the persistence of the
curious mesa-looking like feature in the center of the original ice mound in both
Figures  \ref{GL_1_Exp2_REVISED} and \ref{GL_1_Exp2_QuasiBingham}.  
This is an artifact introduced by the inverted dinner-plate ice distribution. Because
the center of the mound is nearly flat there is hardly any ice movement
there initially.  It is only when the relaxation front that proceeds from the outer
rim works its
way back toward the center of the mound does the flatness of the original distribution
get modified.  The non-circular quality at later times is an expected feature
because the flat region upon which the ice mound was emplaced actually has
some topography in its vicinity (see the corresponding location in the left panel of Fig. \ref{GL-12_InitialBedrock}) 
-- the ice flow that ensues is affected by this relief and, consequently, must move around it
and locally piles up along the local subsurface relief just 
as one expects in classical viscous flow around an obstacle.

\par
\medskip
Figure \ref{GL_2_Composite} shows comparative qualitative results
involving numerical experiments of both basally wet and dry glaciers advancing
on landscape GL-2.  Figures \ref{GL_2_Composite}b-c
show the response of an initial ice mound placed on the sloping highlands
of GL-2, ($x_c = -7$km, $y_c = 7$km) with $\delta r = 4.5$km and
$H\sub{r0} = 0.21$km (Figure \ref{GL_2_Composite}a).  
Figures \ref{GL_2_Composite}b shows the resulting flow shape
when the simulation is run for 19 years in which the glacial ice is dry.
The results are very similar to the developed glacial front 
shown in Fig. \ref{GL_1_LeakyHeights}.  A dry glacier dynamically
evolving with this value of $n=2.1$, which
means it leaks like a fluid down gradient, 
the advancing front weakly exhibits flow lobes upon reaching the flat plains.
\par
Interestingly, Figure \ref{GL_2_Composite}c shows the result of the same
initial configuration except for the difference that the glacial ice is 
allowed to be wet.
In our model, a flowing glacier is considered basally wet (i.e., $q_{sl} \neq 0$), if the 
layer thickness of the ice $H$ exceeds the critical value for melt
at the base which, here, is when $H>H_m = 100$m. This number is chosen to be
consistent with the depth
at which melt might occur if the surface temperatures get as high as 57 \Kelvins
(see discussion in section 3).
Then according to equation (\ref{qslide_def}), the
base of the ice layer will slide downslope with a speed $u\sub{sl0}$, a value which
we have set to $0.2$km/yr, a number adopted from basally wet glacier modeling
for the Earth \cite{Anderson_etal_2012}.
After 19 years, the ice has flown out onto the plains
exhibiting more pronounced flow lobe structure -- qualitatively reminiscent of the lobate features
observed in the New Horizons LORRI image shown in Fig. \ref{Stunning_Glacial_Flow}.
\par
Figure \ref{GL_2_Composite}d shows the result of a similar simulation
except the initial ice profile is centered further up
the landscape ($x_c = -3$km, $y_c = 3$km), with a wider cover
($\delta r = 9.1$km) and with a shallower thickness
($H\sub{r0} = 0.15$km).  The results are similar except that we see
fast moving glacial flow reaching the flats and forming tongue-like
depositional patterns below from several locations.
Also, with a wider initial cover, a siginficant fraction of the ice rapidly infills local topographic
lows. In the particular simulation shown, \N2 ice is still making its
way out of the pitted highlands.

\section{Summary discussion}

In this study we have developed a framework for modeling \N2 glacial ice
and we have used it to answer two broad qualitative questions regarding
the observations of present-day glacial flow observed on the surface
of Pluto.  The glacial model incorporates the known published
thermophysical properties of solid \N2.  The current-day major uncertainty
of the mathematical model
is in the rheological properties of \N2 ice under the extreme conditions
of Pluto's surface.  Once more laboratory and theoretical studies
regarding \N2's rheology are published,
the inputs of the model developed here
will need to be updated accordingly.  However,  the overarching mathematical framework
presented here should remain largely unchanged once updates
to the rheology are determined.
\par
Our glacial flow model assumes the aspect ratio of the flowing ice is very low, which
permits the development of a vertically integrated mathematical formulation (SSA).  
A low aspect ratio means that one may reasonably assume
the ice flows laminarly.
Thus, an important caveat in using {  this thin-layer laminar \N2 flow model }
is the assumption that the layers are not thick enough
for \N2 ice to undergo convection, which would violate the assumption of laminar flow.
{  Questions regarding onset, of course, are subject to the uncertainties in the rheology of \N2
ice -- a matter which needs further laboratory work to clarify.}
\par
As such and as one of the results of this work, we have
analyzed the conditions for the onset of convection for a variety
of {\emph {basal}} thermal boundary conditions and find that it should occur
for layers whose thicknesses are greater than 300-1000m. This figure
that depends upon the \N2 ice grain diameter, whose size distribution is not currently known for
the observed \N2 on Pluto, although upper bounds of 1-10 cms have been
assessed from analysis of spectroscopic data taken with the LEISA instrument \cite{Grundy_etal_2016}.
The thermal boundary conditions considered in our analysis include fixed-temperature
conditions (traditional), fixed-flux, and so-called planetary conditions at the base of the convecting
layer.  The latter two of these boundary conditions, which have not been heretofore
examined in the literature in the context of solid-state convection,
are meant to represent the thermo-physical conditions at the interface of a passively
conducting static bedrock with an overlying \N2 ice layer.  We report that the conditions
for onset when these latter two boundary conditions are assumed
are largely similar to those assessed for the more traditional fixed-temperature basal
conditions \cite{Solomatov_1995,McKinnon_etal_2016}.
\par
The mathematical formulation constructed here also permits the distinction between
basally wet and dry glacial flow.  In a work appearing in this volume (Earle et al., 2017, this volume), obliquity
and orbital precession variations of Pluto can give rise to periods of
time in which Pluto's surface temperature might periodically get as high as 55-60 
\Kelvins \ for 20-30 year stretches of time.  Under such conditions, the base of 
\N2 ice layers as thin as 100-200 meters can melt, which would mean that
one should consider a sliding \N2 layer as a ``basally wet glacier" \cite{Benn_Evans_2010,Anderson_etal_2012}. 
While the physical manifestation of sliding \N2 melt at the base of its solid ice phase
has neither been observed (i.e., either in the laboratory or in images) nor considered
much from a theoretical eye, it is
reasonable to assume that this can happen for Plutonic \N2 ices.  We have both followed
and implemented into our model the formulation for basally wet glacier flow used commonly in the terrestrial 
(water) glaciers
literature, e.g., as found in \cite{Anderson_etal_2012}.  
Given the liquid-solid phase differences between the molecules, we are cautioned
from making firm statements based on the basally wet glacier runs we have reported here,
although we do feel that the gross qualitative trends seen in our simulations
are likely to be robust.
As such, when laboratory investigations relevant to
both wet and basally wet \N2 glaciers are performed in the future, their results
should inform upon improvements to the way in which
such sliding masses are modeled and, moreover, much of the basally wet glacier simulations 
reported here ought be redone. 
\par
The vertically integrated glacial flow has a dry mass-flux form that is similar
to the more familiar Glen-law formulation, which has a power law dependence with
layer thickness, i.e., $\sim H^{n+2}$.  However, based on the strongly
insulating nature of \N2 ice, we find that the mass-flux formulation
deviates from this more traditional form and instead has (what we have here called)
an ``Arrhenius-Glen" form in which the layer thickness dependence
of the mass-flux coefficient is as follows
\[\sim H^{n+2}\exp\left[
\frac{H/H_a}{1+H/H\sub{\Delta T}}\right], \] 
where
the parameters $H_a$ and $H\sub{\Delta T}$ depend upon the activation
energy of \N2 ice, the surface temperature of the glacier, 
and the geothermal flux emerging from the interior of the planet.
We are cautioned by noting that future improvements in the knowledge of \N2 ice rheology
will likely require reconsideration of the mathematical forms and timescales
we have presented in this study.
\par
We have applied this model to qualitatively answer the two
questions we have posed in the introduction:
\begin{enumerate}
\item The wavy transverse light-dark patterning very near the northern
shore of SP (Fig. \ref{Outflow_Image_1}a-b) has been interpreted by us to be evidence of
glacial flow from the center of SP toward the shoreline
(Howard et al., 2017, this volume).  Is the pattern a result of northward advection
downwelling zones of convecting \N2 ice or are they imprints of the bottom
topography present not too far beneath under the surface?  We find
from our simulations that it is likely that the features are steps in the 
flowing \N2 ice tracing out locations of bottom topography.  In many simulations
that we have run, the darkened features correspond to 10-20 m step drops in the ice level
as it passes over the peak of the bottom topography. If the surface ice on Pluto is characterized somehow by a stiffer rheology, perhaps due to contaminants like small \CH4 grains,  
our simulations indicate that the imprint of bottom topography is more dramatic.
Given the currently poor constraints on \N2 ice (and its mixtures), 
we consider this possibility as not being necessarily ruled out.
For stiffer rheologies, the wavy patterns appearing on the surface of the modeled ice
can be an indicator of localized lows in the bottom topography.  In either
case of stiff or non-stiff rheologies, 
the features are ephemeral as they eventually fade away on time scales
approaching a few decades to hundreds of years.  If indeed the
wavy patterning seen near the northern shoreline of SP is an indicator
of bottom topography, then it is reasonable to suggest that this
part of SP has recently experienced a surge of northward flow
within the last two centuries -- perhaps even as recently as in the last few
decades.
\par
\medskip
\item The dark lobate depositional features seen on the eastern shoreline
of SP (Fig. \ref{Stunning_Glacial_Flow}) are another strong line of evidence
pointing at recent glacial flow activity.  Figure \ref{Stunning_Glacial_Flow} 
shows such glacial flow
emanating from the pitted highlands of ETR and onto the flats below.  Were the glacial
flow events responsible for these and similar features a result
of dry or basally wet glacial flow?  Our simulations suggest that it is very hard
to get {\it very pronounced} tongue-like lobate features of down-flowing glacial ice if it moves strictly
as a dry glacier with a power-law index $n$ around 2.  In the simulations we have reported here, the
advancing dry glaciers generally form weakly lobate fronts.
However, if the base of the glacial ice is wet, then it drains out
onto the flats below, exhibiting unmistakably pronounced lobate features
like pointed out in Figs. \ref{GL_2_Composite}c-d.
\par
\medskip
We are not implying that all \N2 glacial ice that drains into SP from
the pitted highlands surrounding SP does so as a wet or basally wet glacier.  However, 
based on the obliquity-eccentricity variations experienced by Pluto,
we consider it perhaps possible that there may have been episodes of rapid
drainage of the \N2 ice found in the pitted highlands of ETR and into SP 
when the surface
temperatures get high enough, since it would mean basal melt 
\N2 ice can occur for shallower layer thicknesses.  
In other words, \N2 ice which would have otherwise
emptied into SP's basin as a dry glacier may have
rapidly drained into the basin as a basally wet glacier during these brief periods
of surface warming.  As a basally wet glacier, then,
they could be the origins of the tongue-like lobate features observed
on today's surface.  {  Of course, we caution that these 
hypotothetical possibilities have not
yet been assessed to their fullest.}
\end{enumerate}
The questions posed and the answers given in this study are very preliminary.
When the rheology of \N2 ice, both as a pure substance as well as an ice comprised
as a mixture of other volatile grains like \CH4 and \CO, most of the simulations and
questions examined here must be revisited.
Similarly, there are many more questions to be answered at this early stage of
discovery.  The qualitative ones we have addressed here give us
confidence that the glacial model we have developed may be used
to answer more detailed and refined questions once more data analysis
of Pluto's surface is complete.  One of the next directions is to apply
the machinery developed here to answer questions about Pluto's past
glaciation (Howard et al., 2017, this volume).

\section{Acknowledgements}
The corresponding author thanks
Magda Saina for assistance with some of the figures presented here. 
The authors thank both the anonymous referee as well as Dr. D. A. Patthoff
for very helpful suggestions that helped to streamline this work.
The authors are also indebted to Dr. W. Grundy for spirited conversations
regarding the granular nature of Pluto's surface ices.
This work was supported by both NASA's Senior NPP program
and NASA's New Horizons project.

\section{References}
\bibliography{mybibfile}

 \appendix
 \section{On the dominance of NH-creep over Coble creep}\label{relative_creep_strengths}
 Volume (NH) and boundary (Coble) diffusional creep mechanisms are (respectively)
 represented in sum
 as
 \beq
 \dot\epsilon = \dot\epsilon_v + \dot\epsilon_b,
 \eeq
 in which
 \beqa
 & & \dot\epsilon_v = \frac{42 D_{0v}\Omega\sigma}{kT d_g^2} e^{-E_v/kT}, \\
 & & 
 \dot\epsilon_b = \frac{42 D_{0b}(2\pi\delta)\Omega\sigma}{kT d_g^3} e^{-E_b/kT},
 \eeqa
 where $E_v,E_b$ are the associated activation energies, 
 $D_{0v},D_{0b}$ are the measured reference diffusion rates,
 $d_g$ is grain size, $\Omega = 4.7\times10^{-29}$m$^3$ is the volume of a single
 \N2 molecule.  The second invariant of the stress-tensor
 is $\sigma$.  In Coble creep the quantity $2\pi\delta$ is a measure
 of the width of the grain boundary through which molecules diffuse and
 this is assumed to be a few widths of a volume of a single \N2 molecule --
 thus we take $\delta \approx \Omega^{1/3}$.
 Laboratory measurements of only volume diffusion parameters (i.e.,
 $D_{0v} = 1.6\times 10^{-7}$m$^2/$s and $E_v = 8.6$kJ/mole)
  are available
 as of the writing of this manuscript.  In lieu of these
 parameters for Coble creep we follow
 \cite{Eluszkiewicz_Stevenson_1990,Eluszkiewicz_1991} who assume that $D_{0b} = D_{0v}$,
 while also assuming $E_b = (2/3)E_b$, the latter being
  a suggestion made originally in
  \cite{Ashby_Verrall_1978}.  The activation energies may be expressed
  in terms of activation temperatures respectively as $T_v \equiv E_v/k = 1033 \ $K
  and  $T_a \equiv E_b/k =  (2/3) E_v/k  = 686 \ $K.
  The point at which the two creep rates are equal defines
  a relationship between temperature and grain size beyond which
  NH creep will dominate over Coble creep.  Thus, setting $\dot\epsilon_v  = \dot\epsilon_b$
  and sorting through the algebra reveals that the critical temperature for this
  transition ($T_{cc}$) occurs when
  \beq
  T_{cc} = T_b \cdot \frac{1}{\ln\Big(d_g/2\pi\Omega^{1/3} \Big)}.
  \eeq
  For a \N2 grain of size 1mm, $T_{cc} = 24$ K.  Thus, given 
  the aforementioned experimental uncertainties, this means under Pluto's current surface
  conditions NH creep is likely to be the dominant operating creep mechanism provided the 
  tangential stresses are well below the power-law regime, a figure
  currently not well constrained ($ < 0.001$ - $1$ MPa).

\section{Infinite Prandtl-number equations}\label{onset_to_convection}
The linearized non-dimensional 2D equations describing the onset of
convection in infinite-Prandtl number media is given by
\beqa
0 &=& -\nabla \Pi + Ra \Theta \hat{\bf z} + \nabla \cdot \nu \nabla {\bf u}
\\
\partial_t \Theta - w &=& \nabla^2 \Theta \\
\partial_x u + \partial_z w &=& 0.
\eeqa
Length and horizontal space dimensions $x$ and $z$ are scaled by the layer thickness $H$,
velocity scales are scaled by $\kappa/H$, while time-scales are given by $\kappa/H^2$.
The quantity $\Theta$ represents the deviations from the static conductive temperature
profile, i.e., $T(z,z,t) = T_s+\Delta T \Theta - \overline T_{_{0z}} z$,
in which $\Delta T = H \overline T_{_{0z}}$.
The Rayleigh number at the base is given as
\beqa
{\rm Ra}
&\equiv&
1.13 \times 10^{13}
\left(\frac{H}{{\rm km}}\right)^4
\left(\frac{d_g}{{\rm mm}}\right)^{-2} \times
\nonumber \\
& & \ \ \ \ \ \ \ \ \  \ \ \   \ \exp\left(
-\frac{T_a}{T_b}
\right)
\left(
\frac{\overline{ T}_{_{0z}}}{{\overline T}_{_{0z}}^{(ref)}}
\right),
\eeqa
where the reference temperature gradient
we adopt is  $\overline T_{_{0z}}^{(ref)}$ = 20 \Kelvins/km,
 the base temperature $T_b \equiv T_s + H {\overline T_{_{0z}}}$
 and the activation temperature $T_a = 1030$ \Kelvins.
 We set the surface position $z_s = 0$.
The function $\nu$ is the viscosity scaled by the viscosity at the bottom of
the layer and is an explicit function of the height parameter $z$
and whose exact form here is
\[
\nu = \exp
\left[
\frac{T_a}{T_s - z {\overline T_{_{0z}}}} - \frac{T_a}{T_b}.
\right]
\]
However, to facilitate calculation we adopt for $\nu$ the Frank-Kamenskii
approximation, e.g. \cite{Solomatov_2012}, which replaces the Arrhenius
form above with a simple exponential, i.e.,
\beq
\nu = e^{pz}.
\eeq
Consequently we identify $\ln p$ as equivalent to the ratio of $\nu_t/\nu_b$, the
ratio of the top viscosity to the bottom viscosity.  This approximation
works well when the viscosity differences become greater than a factor of 1.
A more detailed calculation utilizing the correct Arrhenius form will be reserved
for a future calculation.
\par
We assume a streamfunction formulation for the incompressible perturbation
velocity field, i.e., $u=-\partial_z \psi$ and $w = \partial_x \psi$, 
and we assume steady solutions of the form
\beq
\Theta(x,z) = \Theta_k(z) e^{ikx} + {\rm c.c.}
\eeq
and similarly for the streamfunction $\psi$.  The horizontal
wavenumber is $k\ge 0$ and the vertical eigenfunctions  $\Theta_k(z),\psi_k(z)$ 
are associated with each wavenumber $k$.
\par
For kinematic boundary conditions we employ rigid conditions at the bottom
boundary $z = 0$, $u=w=0$ and stress-free, no-normal flow at $z=1$,
$w = \partial_z u = 0$.  Owing to the fact that the surface of Pluto
today is in vapor pressure equilibrium, we consider it reasonable to impose
fixed temperature boundary conditions at the top.  This means
that $T(z=1) = T_s$, which translates to $\Theta = 0$ at $z=1$.
The bottom thermal condition, on the other hand, is less clear.  If the bottom
ice layer interfaces with an active liquid layer (as in solid-state
convection models of Enceledus
and Europa) then it is reasonable to fix the bottom temperature of the ice layer
to the temperature of the top of the liquid layer, which here effectively
translates to $\Theta = 0$ at $z=0$, and henceforth we refer
to these as {\it fixed-temperature} boundary conditions. 
\par
\begin{figure}
\begin{center}
\leavevmode
\includegraphics[width=9.0cm]{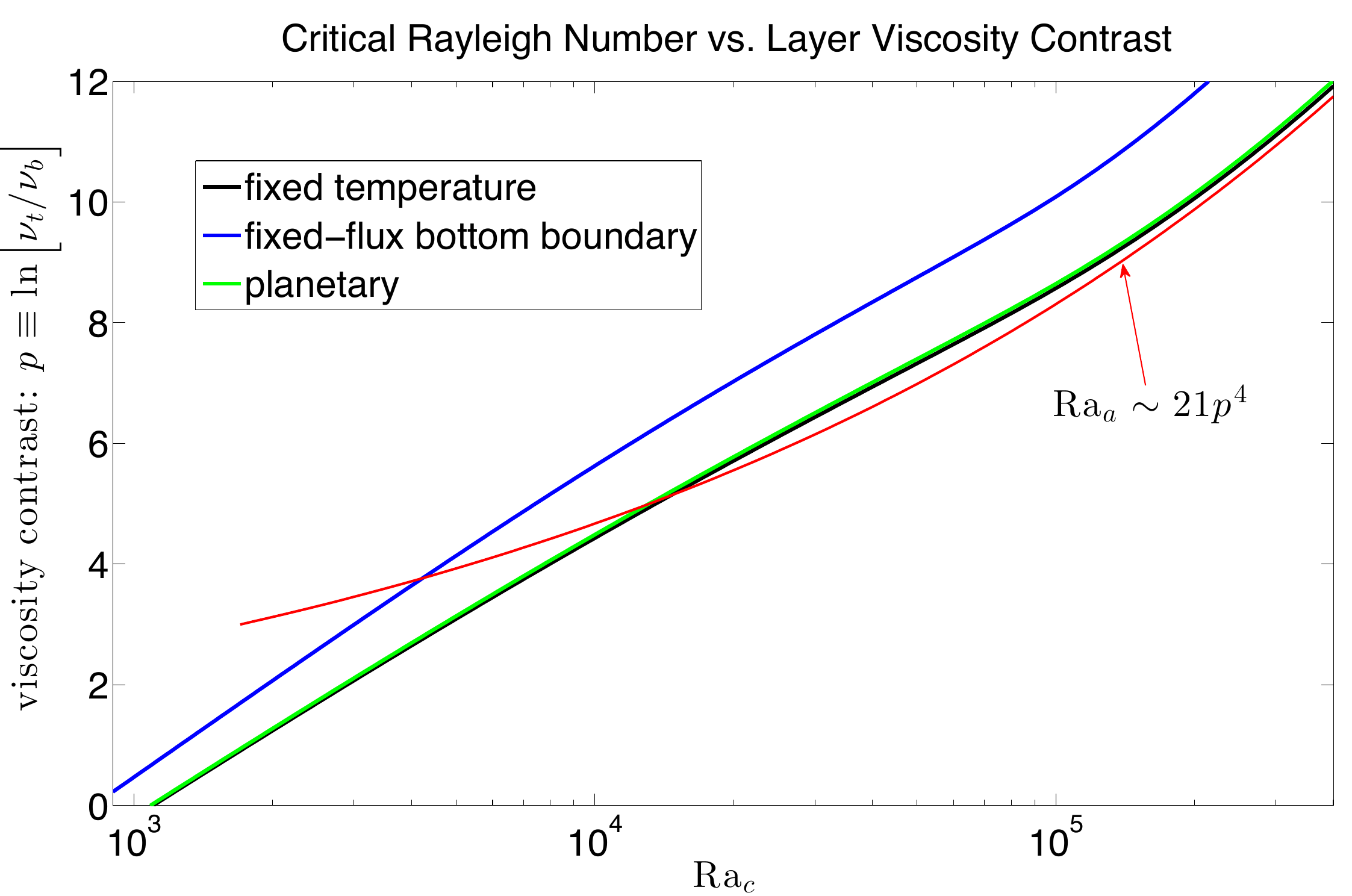}
\end{center}
\caption{Critical value of the basal Ra versus
viscosity contrast $p$ for the onset of solid-state convection
in \N2 ice for parameters shown in Fig. \ref{Min_Layer_thickness_for_Convection}.
Results for three different boundary conditions are displayed.  
The asymptotic relationship Ra$_c \approx \ $Ra$_a = 21 p^4$ is indicated
by the red curve.
}.
\label{critical_Ra_vs_p}
\end{figure}
 However, if the substrate is
a static \water ice ``bedrock", then the correct boundary conditions become
less clear and several options offer themselves here.  The substrate
presumably passively conducts the geothermal flux emanating from the interior
of Pluto and, as such, imposing so-called {\it fixed-flux} boundary conditions
is reasonable, in which $\partial_z T = - {\overline T_{_{0z}}}$ which
translates to $\partial_z \Theta = 0$ at $z=0$.  Perhaps, a more physically accurate
model is to suppose that the \water ice bedrock thermally adjusts
to the temperature fluctuations taking place in the convecting ice-layer
above.  Then following the methodology of \cite{Hurle_Jakeman_Pike_1967}
we match solutions in the convecting ice-layer onto solutions within
the static \water ice bedrock.  As we noted in Section \ref{thermophysical_properties}
the thermal conductivity of \water ice is nearly 20 times that of \N2 ice,
i.e., $K_{H_2O} \approx 20 K_{N_2} $.  Then
according to the procedure outlined in \cite{Sparrow_etal_1964}\cite{Hurle_Jakeman_Pike_1967} the bottom
boundary condition,
 in a calculation examining the conditions of onset (i.e., $\partial_t \rightarrow 0$),
 means imposing
\beq
\partial_z \Theta_k \big|_{z=0} = k \Theta_k\big|_{z=0} (K_{H_2O}/K_{N_2}),
\eeq
for every horizontal wavenumber $k$.
We refer to these thermal conditions as {\it planetary}.
\par
The solution method follows basic procedures, e.g., see \cite{Chandrasekhar_1961},
except for the fact that the critical values of Ra and corresponding eigenfunctions
must be assessed numerically.  We solve the resulting one dimensional sixth order
system on a Chebyshev grid using standard matrix inversion methods found
in all Matlab packages.  We have verified the robustness of the solutions
by comparing the expected critical Ra number for conditions
in which the viscosity is constant.  In the instance when both bottom and
top thermal boundary conditions are fixed-temperature, together with
rigid conditions at the base and stress-free conditions at the top, the
transition Ra = 1101.  Our methods exactly reproduce this value in that case.
\par
We show in Figure \ref{critical_Ra_vs_p} 
the corresponding critical values of the base Ra ($Ra_c$) against the
value of the viscosity contrast $p$ for the three different
thermal boundary conditions described.  We also show the
asymptotic value of Ra$_c \approx \ $Ra$_a = 21 p^4$ which is
 appropriate for temperature dependent Newtonian
convection with fixed-temperature boundary conditions at both boundaries
\cite{Solomatov_1995}.

\section{Glacial Model Development}\label{glacial_flow_model_development}

The details of the glacial flow model is presented here.

\subsection{Modeling framework}\label{modeling_framework}
As we examined in Section \ref{rheological_properties},
glacial ice is assumed to have
a stress strainrate relationship of the following generic form
$
\dot\epsilon = A \sigma ^n,
$
where $A=A(T)$ and $n=n(T)$, and
where $\sigma$ is the local material stress which is a
function of depth from the surface.  
Using Fig. \ref{Modeling_Schematic} as a reference, we see that 
if the gradient of
the local surface slopes with a magnitude angle $\theta$ with respect
to the horizontal, then the corresponding magnitude
of the instantaneous stress is given by
$\sigma = \rho_s g (z_s - z) \tan \theta$ where, as before,
$z_s$ is the height of the local surface located
on the map coordinates $x,y$, i.e. $z_s = Z(x,y,t)$ over time $t$.  
The angle $\theta$ is given by
\beq
\sin\theta = {S}/{\left(1+S^2\right)^{1/2}},
\qquad
S \equiv |\nabla Z| = |\tan \theta|.
\eeq
\par
In the limit of vanishingly small aspect ratios of (nearly) incompressible
flowing materials, it is a fair assumption to approximate
the resulting flow as being purely horizontal with 
no vertical component of velocity.  As such, we assume we are in
a regime in which SSA is valid.
As such, the model procedure in the SSA converts the generic
stress-strain relationship into 
a vertically varying horizontal flow field vector equation,i.e.,
\beqa
 \frac{\partial {\bf u}}{\partial z} &=& A \Big[\rho_s g  (Z-z) \Big]^n (\tan\theta)^{n-1} \ {\bf \nabla} Z \nonumber \\
 &=& A \Big[\rho_s g  (Z-z) \Big]^n {S^{n-1}}
 {\bf \nabla} Z.
 \label{velocity_equation}
\eeqa
where the horizontal velocity
vector is related to its component velocities via
the relationship ${\bf u} = u(x,y,t) {\hat{\bf x}} + v(x,y,t) {\hat{\bf y}}$.
\par

 \begin{figure*}
\begin{center}
\leavevmode
\includegraphics[width=0.9\textwidth]{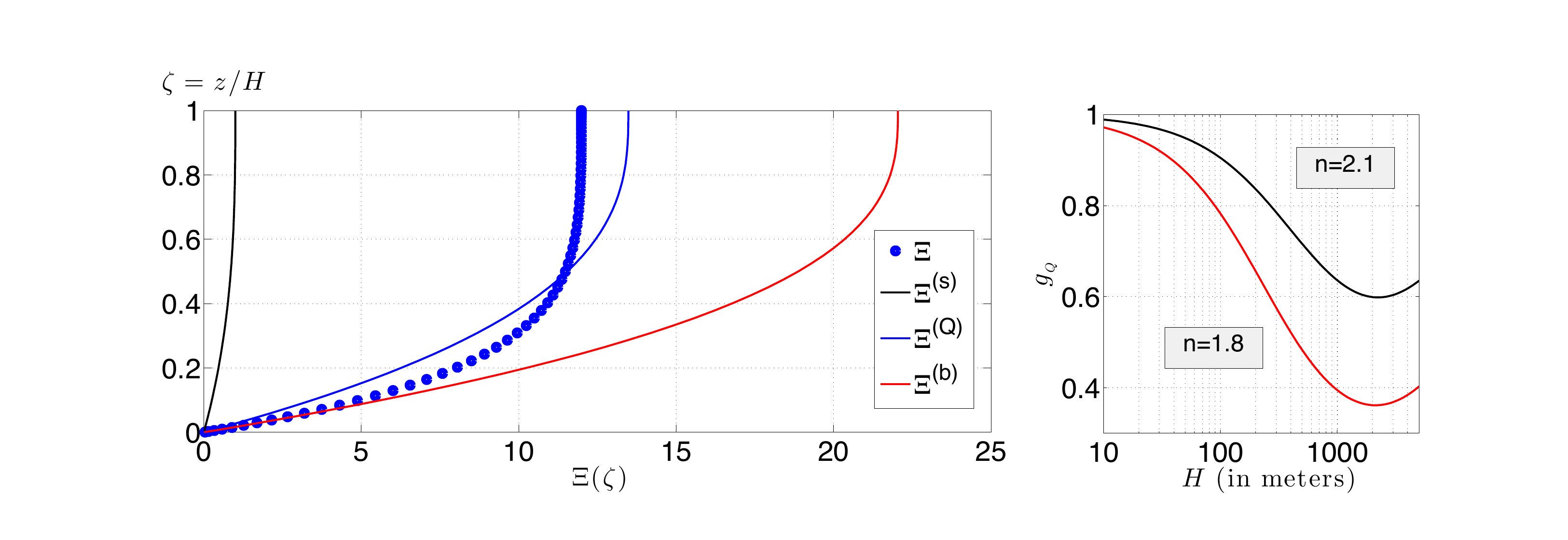}
\end{center}
\caption{Left Panel: Comparison of various approximate forms of $\Xi$.
Figure shows the (scaled) vertical shear profile for a rheology with $n=1.8$,
$T_a = 1036$\Kelvins, $\overline T_{0z} = 20 K/km$, $T_s = 38.5$\Kelvins \ 
and a depth $H=250$m.  Exact solution (filled circles)
compares best with $\Xi^{({\cal Q})}$.
Right Panel: The corrective factor $ g\sub{{\cal Q}}$ for the two rheologies
considered in this paper.  Red curve is appropriate for the \N2 ice-grain rheology of
\cite{Eluszkiewicz_Stevenson_1990}, where $T_a = 1036$\Kelvins, $n = 1.8$, while
the black curve is appropriate for the rheology of laboratory annealed \N2 ice
of \cite{Yamashita_etal_2010}, where  $T_a = 426$\Kelvins, $n = 2.1$.} 
\label{comparative_Xi}
\end{figure*}

The procedure continues by integrating Eq. (\ref{velocity_equation})
once to solve for ${\bf u}(x,y,t)$, subject to a basal boundary condition.
If the ice has not melted (henceforth ``dry"), then we assume rigid boundary conditions
and assign ${\bf u} = 0$
at the base of the layer $z=Z_b(x,y,t) \equiv Z - H$, where $H = H(x,y,t)$
measures the thickness of the ice-layer.  Thus
\beqa
& & {\bf u}(x,y,z,t) 
= {\bf u}\sub{{\rm dry}}(x,y,z,t), \hskip 4.5cm \nonumber
\\
& & {\bf u}\sub{{\rm dry}} \equiv 
\int_{Z_b}^{z}{A \Big[\rho_s g  (Z-z') \Big]^n {S^{n-1}}{}
 {\bf \nabla} Z} dz' \nonumber 
   \eeqa
We often will be interested in diagnosing the surface velocities of
the flowing ice which we define by ${\bf U}(x,y,t) \equiv {\bf u}(x,y,Z,t) $.
\par
Given today's surface temperatures of Pluto ($\sim 38.5$ \Kelvins),
and unless the predominant diameter of \N2 ice grains is much larger than a few 
millimeters,
 it is difficult
to get basal melt of \N2 before the onset of convection inside a layer.
However, as discussed in Earle et al. (2017, this volume) and 
reviewed in section \ref{Pluto_Surface_Temps_Over_Time},
it is possible that the surface temperatures of Pluto's \N2 ices get high enough
so that the base of a \N2 ice layer will exceed its triple point.  While the physical
properties of \N2 ice glaciers with basal melt represents a wholly new subject matter in which
very little
is known both experimentally and theoretically, we consider including this possibility
in our models using very general considerations borrowed from the literature of 
terrestrial \water ice glaciers.  Following \cite{Anderson_etal_2012}, we suppose that
the base of a \N2 ice layer will slide with a constant (prescribed) velocity, 
${\bf u}\sub{{\rm sl}}$
in a direction prescribed by the gradient of the ice surface {\emph{and}} only
 if the 
depth and basal temperatures exceed the minimum required for \N2 melt.  
Thus, we write
\beq
{\bf u}\sub{{\rm sl}}
=
\left\{
\begin{array}{cl}
 u\sub{{\rm sl}0} {\bf {\hat t}}, & H > H_m, \\
{\bf 0}, & 0< H  < H_m, 
\end{array}
\right.
\eeq
in which the map-directional vector ${\bf {\hat t}} \equiv \nabla Z/|\nabla Z|$.
$u\sub{{\rm sl}0}$ is the speed of the basal flow which is a model
input parameter, possibly a function of other inputs.  
The minimum ice-thickness quantity $H_m$ is a model that takes into account
the aforementioned requirements of depth (normal to the ice-layer surface) 
and surface temperature $T_s$.  For the models examined in this study we will
adopt constant values for these quantities, keeping in mind that
matters become far more nuanced and complicated when more physical realism is the aim
\footnote{
There are many shortcomings of this initial model prescription that ought to
be rectified in future modeling.
Some of these issues include the following:
The first, and probably most important, is the way that mixing of melted \N2 is handled.  Liquid 
\N2 is 20 percent more buoyant than its solid phase, and so the matter
of the quality (is it a slush?) and timescale of vertical mixing of liquid \N2 into its solid
phase must be addressed
in a future detailed physical study.  Another matter is the lack of accounting for
the bottom topography at the ice-bedrock interface which can, in fact, hinder or alter the ability
of the basally melted layer to move at all if, for instance, the bedrock shape is locally 
upwardly convex
which would imply the emergence of a trapped pool/lake of liquid \N2
on the bedrock and lying underneath
a covering of solid \N2.  
Until the level
of the liquid phase reaches the brim of the bedrock bowl, one does not expect net flow
due to basal melt.  The model as presented here does not distinguish this subtlety.},
a goal that will be addressed in future follow-up studies.  Nonetheless, in the event that
basal melt happens, then the flow velocity field is  basally wet, which means it
is written as a sum of the dry-field
plus the sliding due to basal melt, i.e.,
\beq
{\bf u} = {\bf u}\sub{{\rm dry}} + {\bf u}\sub{{\rm sl}}.
\eeq
\par
The penultimate stage in developing the model 
involves the formulation of a mass-flux rate ${\bf q}$, defined
as a second integral over the ice-layer of mass flux $\rho_s {\bf u}$, i.e.,
\beq
{\bf Q}(x,y,t) \equiv \int_{z_b}^{Z} \rho_s {\bf u}\ dz
 = 
 Q_0 \nabla Z,
\label{definition_of_mass_flux}
\eeq
where the mass-flux scalar prefactor is 
\[
Q_0 = Q_0(Z_b,H,T_s,T_b,n,\cdots),\] 
in which  $T_b$ is the base temperature
of the ice-layer.  
We explicitly consider the contribution
of the mass-flux due to whether or not basal melt is included.  This is to say that
we identify individual contributions to the mass flux, ${\bf Q}
= {\bf Q}\sub{{\rm sl}} + {\bf Q}\sub{{\rm dry}}
$, as being a sum derived from basal melt and dry flow.  The former
of these is
\beq
{\bf Q}\sub{{\rm sl}} = \rho_s  H {\bf u}\sub{{\rm sl}}
=
\rho_s  H \left\{
\begin{array}{cl}
 u\sub{{\rm sl}0} {\bf {\hat t}}, & H > H_m, \\
{\bf 0}, & 0< H  < H_m, 
\end{array}
\right.
\eeq
Below in subsection \ref{approximate_form_of_dryQ} we explicitly examine the form of the
dry component of the mass flux ${\bf Q}\sub{{\rm dry}}$, as well as the difficulties
it admits.
\par
We add the following ingredients:  We permit the possibility of a steady
form of sublimation/deposition from/onto 
the surface which we denote by $\dot H_a$.  Surface
sublimation/deposition is assumed to occur at a uniform rate {\emph{in the direction
of the local surface normal}} - a formulation used in previous landform evolution
models e.g., \cite{Howard_2007,Howard_Moore_2012}. However we are sensitive to the fact that 
the true amount of deposition/sublimation is a function of the surface
temperatures of the atmosphere, as well as the \N2/CO vapor content of the atmosphere
wherein these quantities are, in turn, functions of PlutoÕs location on its orbit.  Despite these caveats, we
think it fair to consider this more simpler formulation to investigate
the various physical manifestations of basic glacial flow on Pluto's surface.
More sophisticated modeling is reserved for future studies.
\par
Additionally, we permit the possibility that
a certain amount of the substrate bedrock may be eroded by
the moving ice through
the processes of 
scouring or plucking (e.g., \cite{Anderson_etal_2012})
and subsequently absorbed into it and carried along.
\footnote{We realize that this is akin to mixing apples with oranges
as \water ice has a different density and rheology compared to \N2/CO ice.
If the amount of converted substrate bedrock is small compared to the volume
of \N2/CO ice, then we consider the results here to be fair and representative
of the effective volume present of transportable materials.}
We represent this erosion/incision rate by $\dot Z_b$ and allow for the possibility
that it is strictly a function of (i) the total stress exerted at
the glacier/bedrock interface if there is no basal melt, (ii) glacial ice sliding speeds if there is
basal melt and, possibly, (iii) the total ``traffic" of moving ice,
defined as $|{\bf q}|$.  
Thus, the
complete set equations of motion are
\beqa
\partial_t Z &=& \nabla \cdot {\bf q} + \dot H_a + \dot H_b,
\label{evolution_of_Z}
 \\
\partial_t Z_b &=& - \dot H_b,
\label{evolution_of_Zb}
\eeqa
in which ${\bf q} \equiv {\bf Q}/\rho_s$ with
$q_0 \equiv Q_0/\rho_s$ together
where ${\bf Q}$ is given in
Eq. (\ref{definition_of_mass_flux}). 
Because we treat the density as
nearly constant, we prefer to follow
the rate of change of the local elevation $Z$ instead of
the mass-flux rate of change, which is why
we have opted to work with the flux quantity ${\bf q}$.
\par
\bigskip
\subsection{Approximate form for dry ice mass-flux and velocities}\label{approximate_form_of_dryQ}
We examine the form of the velocity fields and corresponding mass-flux in a scenario where there is no basal melt.
In the event that both $A$ and $n$ are independent of depth $z$, then the dry component of
$q_0$ may be evaluated analytically, in which case we say
\beq
q\sub 0 \rightarrow q\sub{{\rm glen}} \equiv
A_c \big(\rho_s g \big)^n \frac{H^{n+2}}{n+2}{S^{n-1}}{},
\label{clean_Q0}
\eeq
recovering the familiar Glen law form \cite{Glen_1955,Benn_Evans_2010,Umurhan_DPS_2015}
.  Also in this limiting case we can
exactly assess both the horizontal velocity field of the dry \N2 ice,
\beqa
 {\bf u}(x,y,z,t) \rightarrow  {\bf u}^{(c)}(x,y,z,t) =  
 \hskip 3.0cm & & \nonumber \\ 
 \frac{A_c H}{n+1}\big(\rho_s g H\big)^n
 \left[{1 - \left(\frac{Z-z}{H}\right)^{n+1}}\right]
{S^{n-1} \nabla Z}{}, & & 
\label{clean_u}
\eeqa
and similarly the surface velocity of the dry flowing ice, i.e.,
\beqa
 {\bf U}(x,y,t) \rightarrow  {\bf U}^{(c)}(x,y,t)
= \ \ \ \ \ \ \ \ \ \ \ \ \ \ \ \   & & \nonumber \\
\frac{A_c H}{n+1}\big(\rho_s g H\big)^n {S^{n-1}\nabla Z}{}
, & & 
\label{clean_U}
\eeqa
where the superscript ``(c)" denotes the corresponding expression
for constant values of $A = A_c$ and $n$.  For the following comparative
analysis,
it is useful to rewrite the velocity field as
as a product of two functions
\beqa
{\bf u}^{(c)}(x,y,z,t) &=& {\bf U}^{(c)}(x,y,t) \cdot \Xi^{(c)}(z), 
\label{u_c_def} 
\\
 \Xi^{(c)}(z) & \equiv & 
{1 - \left(\frac{Z-z}{H}\right)^{n+1}}.
\label{Xi_c_def} 
\eeqa
 \par
 \N2 ice rheologies have temperature dependencies, namely
 in the pre-factor $A$ and stress-strain exponent $n$.  Given
 our earlier remarks in Sections \ref{thermal_properties} 
 and \ref{convective_onset},  
 and assuming the ice-layers under consideration
 are not undergoing convection, the interior
 temperatures increase linearly with depth which
 makes writing down clean formulae for $Q_0$, ${\bf u}$
 and ${\bf U}$, as in Eqs. (\ref{clean_Q0}-\ref{clean_U}),
 unwieldy in most cases --
 especially when $A$ is characterized as an Arhenius function
 of temperature.  Nonetheless, to clarify matters and
 help motivate a useful approximate form, we express
 the exact solution for ${\bf u}$ as a similar product of
 two functions, i.e.
 \beq
 {\bf u} \equiv {\bf U}^{(c)} \Xi(z),
 \eeq
 where we evaluate $A_c \rightarrow A_s = A(T=T_s) =  A(z=Z)$ and
 letting
 \beqa
 & & \Xi(z) \equiv (n+1)\int_{\zeta_b}^\zeta \frac{A(\zeta')}{A_s}\left(\frac{Z}{H}
 - \zeta'\right) d\zeta', 
 \label{def_exactXi}
 \\
 & & \ \ \ \zeta \equiv z/H, \qquad \zeta_b \equiv Z_b/H.
 \eeqa
 Our goal then is to develop
 an approximate solution, $\Xi^{(appx)}$, for the exact vertical
 structure function $\Xi(z)$ -- the latter of which must be determined
 numerically.

 \par
 We are reminded that the index $n$ weakly varies with respect
 to temperature
 based on the laboratory annealed \N2 experiments of \cite{Yamashita_etal_2010},
 viz. Section \ref{rheological_properties}, and that
 the strongest variations will arise due to the
 temperature dependence of $A$.
 A practical strategy then to handle this is to adopt a constant
 value of $n$ and consider two end-state values of $A$: one
 corresponding to its value based on the surface temperature
 and one based on the temperature at the base of the layer.  In other
 words, one proceeds by adopting the 
 forms for $Q_0$, ${\bf u}$
 and ${\bf U}$ given in Eqs. (\ref{clean_Q0}-\ref{clean_U})
 but
 replacing $A_c$ with $A_s = A(T_s)$ for the upper end-state
 and the other lower-end state by replacing $A$ with $A_b = A(T_b)$
 where $T_b$ is the bedrock temperature
 as discussed in Section \ref{convective_onset}.
 We consider two choices for the approximate vertical structure
 function form $\Xi^{(appx)}$: one is where one 
 sets $A(\zeta') \rightarrow A_s$ in the definition found
 in Eq. (\ref{def_exactXi}), i.e.
 \beq
 \Xi^{(s)} = 
 (n+1)\int_{\zeta_b}^\zeta \left(\frac{Z}{H}
 - \zeta'\right) d\zeta' = \Xi^{(c)},
 \eeq
 and the second one is
 $A(\zeta') \rightarrow A_b$
 \beqa
  & & \Xi^{(b)} = 
 (n+1)(A_b/A_s) \int_{\zeta_b}^\zeta \left(\frac{Z}{H}
 - \zeta'\right) d\zeta' \nonumber \\
 & &  \ \ \ \ \ \ \ \ \  = \exp\left[\frac{ H/H_a}{1 + H/H\sub{\Delta T}}\right] \Xi^{(c)},
 \eeqa
  in which
  the ratio $A_b/A_s$ was evaluated assuming the generic Arrhenius
  functional forms posited in Section \ref{rheological_properties}.
  The quantities appearing are $H\sub{\Delta T} \equiv T_s/\overline T\sub{0z}$
 and $H_a \equiv T_s^2/T_a \overline T\sub{0z} = H\sub{\Delta T}\cdot(T_s/T_a)$.
 We consider a third possibility, which is similar to $\Xi^{(b)}$ except it
 is multiplied by a corrective factor $g\sub{{\cal Q}}$ chosen
 to produce an exact scalar equivalence
 between the exact mass flux and this approximate form.  Thus, we also
 consider
  \beqa
  & & \Xi^{({\cal Q})} = g\sub{{\cal Q}}  \Xi^{(b)} \nonumber \\
 & &  {\displaystyle
  g\sub{{\cal Q}} = \int_{\zeta_b}^{Z/H} \Xi(\zeta) d\zeta
   \Bigg/ \int_{\zeta_b}^{Z/H}  \Xi^{(b)}(\zeta) d\zeta.
  }
 \eeqa
 As developed, we have three approximate
 forms of the velocity field ${\bf u}$, i.e.
 ${\bf u} \approx {\bf U}^{(c)}\Xi^{(s)}, \ 
 {\bf u} \approx {\bf U}^{(c)}\Xi^{(b)}$
 and $ {\bf u} \approx {\bf U}^{(c)}\Xi^{({\cal Q})}$,
 and we must now compare these to the exact numerically determined
 form, ${\bf u} = {\bf U}^{(c)} \Xi$.
 Because the horizontal functions ${\bf U}^{(c)}$ are
 the same across all test functions, it will be enough
 for us to assess the robustness of a given approximation
 by comparing the three vertical functions
 $\Xi^{(s)}, \Xi^{(b)}$ and $\Xi^{({\cal Q})}$ 
 against the numerically determined exact function $\Xi$.
 The left panel of Fig. \ref{comparative_Xi} shows the vertical
 structure functions discussed here and as compared to the
 exact solution determined using a Chebyshev integration method.
 We see that the exact solution $\Xi$ is well approximated
 by $\Xi^{({\cal Q})}$.  The right panel of the same
 figure also proves that $\Xi^{({\cal Q})}$ is
 also well approximated by $ \Xi^{(b)}$, the use of the latter
 which would incur errors of no more than 50\%.  The implication
 of this is that the flow rates and mass-fluxes are reasonably
 well approximated by a factor of 2 (i.e., $\sim$max$(1/g\sub Q)$)
 when utilizing the rates associated
 with the higher temperatures corresponding to the base of the layer,
 i.e., by approximating the velocity field
 with end-state defined by the
 base of the layer, ${\bf u} \approx {\bf U}^{(c)} \Xi^{(b)}$.
 
  Consequently, and from here on out, we adopt the approximate
 dry ice velocity field constructed from $\Xi^{({\cal Q})}$ 
 and explicitly write out the analytical expressions for
 the dry components of the \N2 ice mass-flux and velocities
 to be used
  in our numerical modeling hereafter:
 Beginning with the surface dry velocities we have,
 \beqa
 & & {\bf U}(x,y,t) \approx  g\sub{{\cal Q}}
 \exp\left[\frac{ H/H_a}{1 + H/H\sub{\Delta T}}\right] {\bf U}^{(c)}(x,y,t) 
 \nonumber
 \\
 & &   =
 g\sub{{\cal Q}}
 \exp\left[\frac{ H/H_a}{1 + H/H\sub{\Delta T}}\right]
 \frac{A_s H}{n+1}\big(\rho_s g H\big)^n {S^{n-1}\nabla Z}{}, \nonumber \\
 & & 
 \label{U_apprx_def}
 \eeqa
 whereupon, the vertically sheared total horizontal flow (dry plus sliding fields) becomes
 \beq
 {\bf u}(x,y,z,t) \approx {\bf u}_{{\rm sl}} + {\bf U}(x,y,t) \Xi^{(c)}(z),
 \label{u_appx_def}
 \eeq
 where $\Xi^{(c)}$ is as given in Eq. (\ref{Xi_c_def}).


\end{document}